\documentclass[a4paper,11pt]{article}
\usepackage[margin=1in]{geometry} 

\usepackage[numbers]{natbib}

\usepackage[ruled,vlined,linesnumbered]{algorithm2e}
\SetAlFnt{\footnotesize} 
\SetAlCapFnt{\footnotesize}
\SetAlCapNameFnt{\footnotesize}
\usepackage{array}
\usepackage{textcomp}
\usepackage{stfloats}
\usepackage{url}
\usepackage{verbatim}
\usepackage{graphicx}

\usepackage{amssymb}
\usepackage{ifthen}
\usepackage[utf8]{inputenc}
\usepackage{latexsym, dsfont}
\usepackage{amsmath,amsthm}
\usepackage{comment}
\usepackage{siunitx}
\usepackage{graphicx}
\usepackage{caption}
\usepackage{hyperref}   
\usepackage[capitalize]{cleveref}
\usepackage{todonotes}
\usepackage[caption=false,font=footnotesize]{subfig}
\usepackage{mathtools}
\usepackage{tikz}
\newboolean{PGFPlots}
\setboolean{PGFPlots}{true}

\newtheorem{thm}{Theorem}[section]

\newtheorem{prp}[thm]{Proposition}

\newcommand{\dup}{\delta_\uparrow}
\newcommand{\ddo}{\delta_\downarrow}

\newcommand{\dupD}{\delta^\uparrow}
\newcommand{\ddoD}{\delta^\downarrow}

\newcommand{\dsu}{\delta_S^\uparrow}
\newcommand{\dsd}{\delta_S^\downarrow}

\newcommand{\figPath}[1]{figures/#1}

\newcommand{\NOR}{\texttt{NOR}}
\newcommand{\NAND}{\texttt{NAND}}

\newcommand{\cg}{\texttt{C}}
\newcommand{\invt}{\texttt{InvTool}}
\newcommand{\AOIgate}{\texttt{AOI}}

\newcommand{\dmin}{\delta_{\mathrm{min}}}

\newcommand{\hspice}{\textit{HSPICE}}
\newcommand{\spice}{\textit{SPICE}}

\newcommand{\vth}{V_{th}}

\newcommand{\vdd}{V_{DD}}

\newcommand{\dd}{\mathrm{d}}

\newcommand{\vint}{V_{int}}

\newcommand{\nmos}{\texttt{nMOS}}
\newcommand{\pmos}{\texttt{pMOS}}

\newcommand{\ohm}{(OHM)}

\newcommand{\on}{\mbox{\emph{on}}}
\newcommand{\off}{\mbox{\emph{off}}}

\newcommand{\bA}{\bar{A}}
\newcommand{\bchi}{\bar{\chi}}
\newcommand{\bd}{\bar{d}}
\newcommand{\bc}{\bar{c}}

\newboolean{conference}
\setboolean{conference}{true}  

\begin{document}

\title{Drafting and Multi-Input Switching in Digital Dynamic Timing Simulation for Multi-Input Gates\thanks{This is an extended version of the paper accepted and published at the Design, Automation and Test in Europe Conference (DATE'26).}}

\author{
  Arman Ferdowsi$^{1,}$\thanks{Corresponding author} \and
  Ulrich Schmid$^{2}$ \and
  Josef Salzmann$^{3}$
}

\date{
  $^{1}$University of Vienna, Faculty of Computer Science\\
  $^{2}$TU Wien, Institute of Computer Engineering, Embedded Computing Systems Group\\
  $^{3}$TU Wien, Institute for Microelectronics\\
  \emph{arman.ferdowsi@univie.ac.at} \quad \emph{s@ecs.tuwien.ac.at} \quad \emph{josef.salzman@tuwien.ac.at}
}

\maketitle

\begin{abstract}
\noindent
We present a prototype multi-input gate extension of the publicly available Involution Tool for accurate digital timing simulation 
and power analysis of integrated circuits introduced by \"Ohlinger et al.\ (Integration, 2021). Relying on discrete event simulation, the Involution Tool allows 
fast timing simulation of circuits made up of an arbitrary 
composition of supported gates, provides automatic random input stimulus generation,
and supports parameter sweeping. It also enables a detailed comparison of the delay predictions obtained by
different models, including pure and inertial delays as well as digitized SPICE-generated 
reference traces. Our extension added support for 2-input gates like NOR and NAND, by implementing 
novel analytic delay formulas obtained via a refined analysis of a recently proposed thresholded 
first-order hybrid model of such gates. The resulting formulas faithfully cover not only multi-input switching 
effects (also known as Charlie effects), but also the decay of short pulses (aka Drafting effects).
Besides the fact that our analytic models not only allow the derivation of closed-form delay
formulas for arbitrary compositions of such gates, they are also key for a strikingly simple
procedure for model parametrization, i.e., for gate characterization, which only needs three
characteristic delay values.
Using the extended Involution Tool, we compare the delay and power predictions for some
benchmarking circuits stimulated by randomly generated input traces. Overall, our results reveal considerably improved prediction accuracy compared to the original Involution Tool, without a 
noticeable performance penalty.
\end{abstract}

%


\section{Introduction}
\label{sec:intro}

Fast and accurate timing analysis techniques are crucial for modern digital circuit design. State-of-the-art \emph{static timing analysis} (STA) techniques, e.g.\ based on CCSM~\cite{Syn:CCSM} and ECSM~\cite{Cad:ECSM} models, facilitate fast and accurate corner case analysis of worst-case and best-critical path delays even in very large designs. However, STA can deal with PVT variations and dynamic effects like slew-rate variations, crosstalk, and 
\emph{multi-input switching} (MIS) effects \cite{CS96:DAC,CGB01:DAC} only via appropriately increasing the 
margins of the delay estimates, see \cref{sec:STA} for a survey of such approaches. \emph{Statistical} static timing analysis (SSTA) techniques \cite{BCSS08:TCAD,FP09:VLSI} have been developed to mitigate this problem: Relying on some statistical model of the variabilities, SSTA approaches like \cite{ADB04:DAC,SZ05:ICCAD,YLW05,FTO08:DAC,TZBM11,SRPW20:DAC} are used to compute statistical delay estimates. Whereas this works very well for effects like process variations,
where accurate statistical models can be provided, it is not suitable for effects that depend on the 
actual trace of the signals generated by a circuit.

Indeed, the delay estimates provided by (S)STA do not take into account 
the \emph{actual} trace history for a given signal transition, but only
the worst-case resp.\ best-case one. The delay of a gate stimulated by 
this particular transition may depend strongly on its history, however: Even
for a single-input-single-output gate like an inverter, an input
transition causing the output to change shortly after the previous
output change will have a shorter delay than an output change that happens
only after the previous one has saturated (this variability has been called the
\emph{drafting effect} in \cite{CharlieEffect}). For multi-input gates,
gate delays may also vary when transitions at \emph{different} inputs
happen in close proximity (such MIS effects are also known as
\emph{Charlie effects}, named after Charles Molnar, who identified
their causes in the 1970).

Consequently, in the case of a timing violation reported by STA
in some critical part of a circuit, which may in fact be a false positive
originating in excessive margins, verifying the correct operation requires a detailed analysis
of the actual signal traces generated by the circuit. Facilitating this is the realm 
of \emph{dynamic timing analysis} (DTA) techniques. Their necessity
can be nicely exemplified by means of the token-passing ring described
and analyzed 
by Winstanley et al. in \cite{CharlieEffect}, for example: This asynchronous
circuit implements a ring oscillator made up of stages consisting of a
2-input Muller \cg\ gate, the inputs of which are connected to the previous resp.\ next
stage. The authors uncovered that the ring exhibits two 
modes of operation, namely, burst behavior versus evenly spaced output transitions, 
which can even alternate over time.
The actual mode depends on some subtle interplay between the drafting
effect and the Charlie effect of the \cg\ gates in the circuit. 
To understand the behavior of this circuit, \emph{individual} transitions must hence be traced 
throughout the ring.

Analog simulations, like SPICE \cite{NP73:spice}, are the gold standard for DTA. Since they require the numerical solution of the system of \emph{ordinary differential equations} (ODEs) used for describing the behavior of the transistors, analog simulations are prohibitively time-consuming even for small circuits and short signal traces, however. \emph{Digital} DTA (DDTA) techniques, which rely on \emph{delay models} that allow one to predict gate delays on a per-transition basis, have been proposed as a less accurate but much faster alternative here. State-of-the-art tools like Mentor Graphics ModelSim/QuestaSim, Cadence NC-Sim or Synopsys VCS use models like CCSM or ECSM for parametrizing pure or inertial delay channels \cite{Ung71}, e.g., in VHDL Vital or Verilog timing libraries, and then use the resulting executable HDL simulation models in subsequent simulation and timing analysis runs. This allows fast functional verification and reasonably accurate performance and power estimation for large circuits already early in the development cycle \cite{najm1994survey,FB95}.

However, whereas inertial delay models capture the suppression of short pulses, neither pure nor inertial delay models 
can model pulse degradation, as it occurs when a very short input pulse is fed into a buffer or inverter, for example, see \cref{Sec:MIS} for more details. 
Pulse degradation can be modeled by means of \emph{single-history} delay models as introduced in \cite{FNS16:ToC}, however, where a gate's input-to-output delay $\delta(T)$ may depend on the previous-output-to-input delay $T$. Note that this dependency on $T$ captures the drafting effect \cite{CharlieEffect} mentioned above. Unfortunately,
the state-of-the-art here (see \cref{sec:DDTA}) is much less developed than for STA (see \cref{sec:STA}):
The only instances of single-history models known to us are the \emph{delay degradation model} (DDM) \cite{BDJCAVH00,BJV06}, which uses an exponential function for $\delta(T)$, and the \emph{involution delay model} (IDM) \cite{FNNS19:TCAD}, the delay function of which is a negative involution $-\delta(-\delta(T))=T$. 

The present paper builds on a recent extension \cite{ferdowsi2024hybrid} of the IDM, which provides delay
models for interconnected 2-input \NOR\ and Muller \cg\ gates that cover all MIS effects accurately: In the case of a 2-input \NOR\ gate, 
for example, the input-to-output delay ($\dup(\Delta)$ for rising output transitions and $\ddo(\Delta)$ for falling ones)
varies with the switching time difference $\Delta$ of \emph{different} inputs, see \cref{Sec:MIS} for more information. One of the main benefits of the
availability of such \emph{analytic} delay formulas is a parametrization procedure,
which allows to compute all model parameters by computing functions that
only require three characteristic MIS delay values of the to-be-characterized
  gate ($\delta(0)$, $\delta(\infty)$ and $\delta(-\infty)$, and some uncritical estimate of
  its waveform-independent pure delay $\dmin$) as their input. Since these delay
  values could be easily incorporated in a gate library, the usually tedious
  process of gate characterization becomes extremely simple and fast.

  However, in \cite{ferdowsi2024hybrid}, the additional influence
  of drafting effects on gate delays has not been considered. 
  To support the needs of accurate DDTA, however, a proper delay model
  a 2-input \NOR\ gate should provide delay functions 
$\dup(T,\Delta)$ and $\ddo(T,\Delta)$, which vary with both $\Delta$ and $T$. Developing such delay formulas, implementing those in the Involution Tool,
and experimentally evaluating their prediction accuracy, both by comparison to SPICE-generated data and by
performing DDTA of some representative circuits, is the main purpose of this paper. 

\medskip

\textbf{Main contributions:}
\begin{enumerate}
\item We build on the analytic results of \cite{ferdowsi2024hybrid} to develop a complete 
set of analytic delay functions $\dup(T,\Delta)$ and $\ddo(T,\Delta)$ for a 2-input 
\NOR\ gate, for arbitrary choices of $\Delta$ and $T$. We also show how derive the resulting
formulas for a 2-input \NAND\ gate from those.


\item We provide a simulation algorithm, which, given a sequence of input transitions, computes 
the corresponding sequence of output transitions that paves the way for processing digital signal propagation throughout the gate.


\item We extend the Involution Tool by implementing our simulation algorithm and our extended
  delay models for \NOR\
and the ``dual'' \NAND\ gates, and briefly describe the resulting features and some details of the implementation.

\item We conduct some simulation experiments on two representative benchmarking circuits,
  including the $\NOR$-eqivalent of the \texttt{c\_17\_slack} taken from \cite{ISCAS85_reference}
  already used in \cite{OMFS20:INTEGRATION}, and evaluate the achievable accuracy, by comparing
  its timing predictions to alternative approaches.
\end{enumerate}

Whereas we consider our improvement of the state-of-the-art of accurate simulation-based DDTA significant, we want to stress that the usage of our closed-form analytic delay formulas
is not restricted to simulation-based approaches: 
They can seamlessly be integrated into state-of-the-art formal verification 
frameworks like \cite{AH24} as well, thereby enabling automatic end-to-end 
correctness proofs under realistic timing assumptions. Even more, they unlock 
the development of novel symbolic timing analysis approaches that go far beyond 
existing approaches like \cite{CC04}. Indeed, part of our ongoing research 
\cite{TEFS25:arxiv} is devoted to a computer-algebra based tool,
which allows to compute delay formulas for a compound circuit by analytically 
composing the delay formulas of the constituent gates, and to \emph{analytically} 
determine the dependency of certain critical delays on individual
gate parameters. Note that this can be viewed as a way to carry over and further
generalize \emph{differentiable} timing analysis approaches, 
which are being used in recent static timing analysis techniques like NVIDIA's 
\emph{INSTA} framework \cite{luinsta}, to dynamic timing analysis.

\textbf{Paper organization:}
In \cref{sec:relwork}, we briefly survey related work, both in the context of STA and DDTA,
in \cref{Sec:MIS}, we informally explain drafting and MIS effects in multi-input gates. \cref{sec:Prel} is
devoted to a brief recap of the cornerstones and properties of the thresholded hybrid model for 
an interconnect-augmented \NOR\ gate 
proposed in \cite{ferdowsi2024hybrid}. Building on these results, we develop a 
fully-fledged digital delay model in \cref{Sec:General_Delay_Model}, which comprises analytic 
formulas for $\dup(T,\Delta)$ and $\ddo(T,\Delta)$ and the simulation algorithm for computing output transition 
sequences. In \cref{sec:idm_sim}, we revisit the main features of the existing Involution Tool and 
describe our extensions. \cref{Sec:experiments} describes our experiments and their results. Some
conclusions in \cref{Sec:con} round off our paper.

\section{Related Work}
\label{sec:relwork}

In this section, we briefly report on related work devoted to delay models 
for static timing analysis (\cref{sec:STA}) and accurate digital 
dynamic timing analysis (\cref{sec:DDTA}).

\subsection{Delay models for STA}
\label{sec:STA}

A number of delay models for STA have
been proposed in the literature, which also capture MIS effects. 
In~\cite{CS96:DAC}, Chandramouli and Sakallah resorted to
macromodels, i.e., blackbox functions involving delay-relevant input
parameters like load capacitance as well as the input transition time(s),
which compute the gate delay. They also show how to compose 2-input macromodels
to get $n$-input macromodels. Their approach achieves a delay and slew rate
prediction accuracy in the $1 \dots 10$\% range.

In \cite{CGB01:DAC}, the authors develop empirical delay formulas that
cover MIS effects for 2-input gates, and use curve fitting (based on detailed
simulation data, for every gate) for determining the appropriate parameters. 
They also show how to incorporate their model into STA, using 
incremental timing refinement (ITR) to reduce the margins. 
A similar approach has been proposed in~\cite{SKJPC09:ISOCC}, where
a quadratic polynomial is used for the delay formula. In \cite{SRC15:TODAES},
Subramaniam, Roveda, and Cao used a piecewise linear delay function for
this purpose.
The authors use propagation windows to reduce the margins resulting
from using their modeling approach in STA.

A machine-learning-based MIS modeling approach has been proposed
in~\cite{RS21:TCAD}. Whereas it is distinguished by its very
good accuracy, which is in the few \%-range, its downside is the
inevitable per-gate training requirement.


All the MIS models surveyed above target the delay functions 
themselves. A very different type of delay model is obtained
by developing a simplified model of a gate and determining the 
resulting delay function. In the approach proposed in \cite{AKMK06:DAC},
Amin et~al.\ used an analog model that consists of suitably
defined non-linear resistors and capacitors at each pin of 
a gate. Circuits are composed by composing the appropriate
models of the gates along the circuit's paths. Accuracy validation 
experiments revealed an accuracy in the $1 \dots 10$\% range.

\subsection{Delay models for accurate DDTA}
\label{sec:DDTA}

\begin{figure}[h]
  \centerline{
    \includegraphics[width=0.55\linewidth]{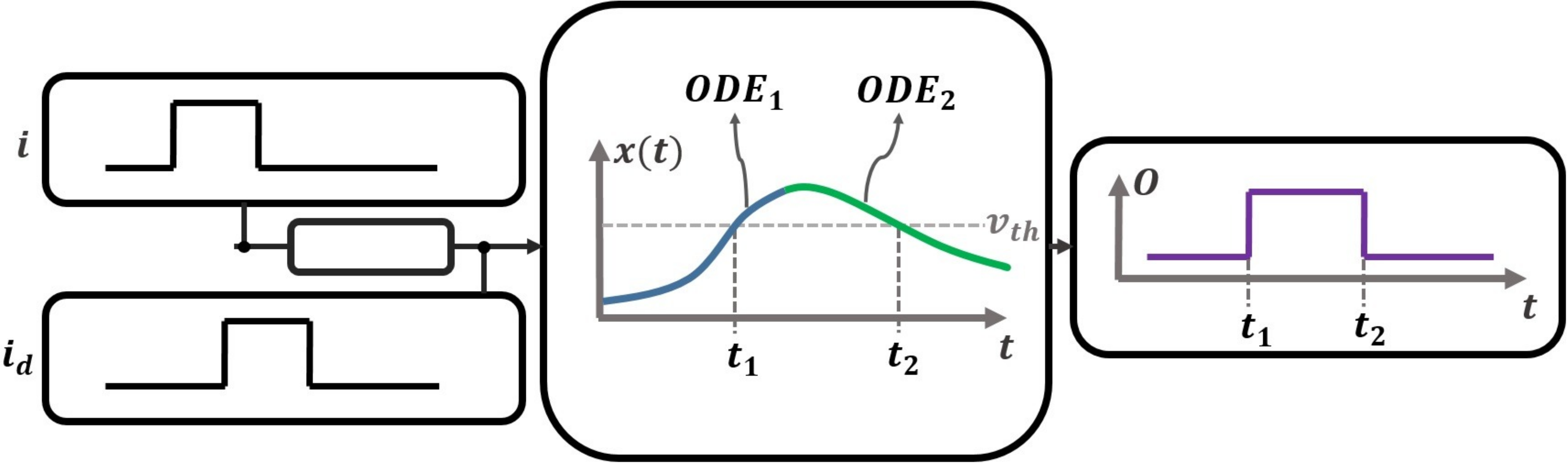}
  }
  \caption{Illustration of the thresholded hybrid system of an IDM channel, with a single input $i$ and output $o$.
  It comprises an (optional) pure delay shifter, producing $i_d$, and two ODEs governing some state signal $x(t)$
  that is digitized by a threshold voltage comparator to produce $o$. The active ODE is selected by the current 
  state of $i_d$, with mode switches that guarantee continuity of $x(t)$.}
  \label{fig:switched}
\end{figure}

In order to capture the drafting effect, digital delay models suitable for DDTA must at least be single-history \cite{FNS16:ToC}. Among all known delay models \cite{Ung71,BDJCAVH00,BJV06,FNNS19:TCAD} for DDTA, the IDM is the only one that faithfully models glitch propagation in the canonical short-pulse filtration problem, as defined in \cite{FNS16:ToC}. Another particularly compelling feature of the IDM is its
simplicity, which originates in the fact that it can be viewed as a simple 2-state thresholded first-order hybrid model \cite{FFNS23:HSCC}. As illustrated in \cref{fig:switched}, in such a system, the state of the digital input is used to select one among two \emph{ordinary differential equations} (ODEs) that govern some internal analog signal, the 
digitized version of which constitutes the digital output. This simplicity allows one to derive explicit analytic formulas for the IDM channel delays $\delta(T)$, which are instrumental for very fast digital timing simulation.

The IDM also comes with a publicly available discrete-event simulation framework,
the \emph{Involution Tool} \cite{OMFS20:INTEGRATION}. It also allows to compare IDM delay predictions against SPICE-generated data and other delay models, for circuits composed of elementary gates supported by the tool. The Involution Tool has been used for demonstrating very good accuracy of the IDM for inverter chains and clock trees, achieving a 250-fold speedup in simulation times compared to SPICE. For circuits also incorporating multi-input gates like \NOR, however, the achievable accuracy of the IDM degrades. The authors of \cite{OMFS20:INTEGRATION} conjectured that this is primarily a consequence of the fact that the simplistic single-input-single-output channel model of the original IDM cannot model MIS effects. 

In an attempt to mitigate this problem, Ferdowsi et~al.\ developed \emph{thresholded hybrid models} 
for 2-input CMOS \NOR\ and \NAND\ gates \cite{FMOS22:DATE, ferdowsi2025faithful}. These models 
are based on replacing transistors by (possibly time-varying) switched resistors and are all faithful. 
It turned
out that all the MIS effects are covered by the model proposed in \cite{ferdowsi2025faithful},
which is a 4-state hybrid first-order model inspired by the Shichman-Hodges transistor model~\cite{ShichmanHodges}. 
Whereas the ODEs governing its 4 modes are all first-order, they have time-varying coefficients and are 
hence not trivial to solve analytically. Nevertheless, the authors finally managed to develop simple 
and accurate analytic delay formulas for this model. Moreover, thanks to a simple and fast parametrization procedure, only the extremal MIS delay
values ($\delta(0)$, $\delta(\infty)$ and $\delta(-\infty)$) are needed for computing the model
parameters for a given gate. 

In \cite{ferdowsi2024hybrid}, both the model and the parametrization procedure developed in \cite{ferdowsi2025faithful} for ``naked'' gates was extended to gates with an interconnect. 
Using a comparison to SPICE-generated data, the authors showed
that appropriately parametrized versions of their model predict the actual delay of 2-input \NOR\ and Muller \cg\ gates implemented in different CMOS technologies and a for wide range of different interconnects 
accurately and very fast, for any value of the input separation time $\Delta$.
The drafting effects has not been considered in \cite{ferdowsi2024hybrid}, however.

\section{Multi-Input Switching and Drafting}
\label{Sec:MIS}

\begin{figure}[ht]
  \centerline{
    \includegraphics[width=0.7\columnwidth]{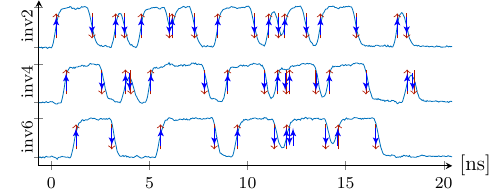}
  }
  \caption[Waveform UMC-90]{Measured waveform for a $90$nm CMOS inverter chain, with
    the predictions according to the IDM (red long up/down-arrows) 
and the DDM (blue short up/down arrows). Taken from \cite[Fig.~8]{FNNS19:TCAD}.}
  \label{fig:wave90_inv}
\end{figure}

\begin{figure}
  \centerline{
    \includegraphics{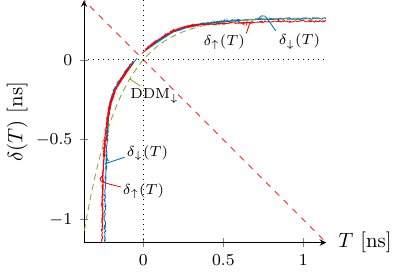}}
  \caption{Measured $\ddo$ (blue) and~$\dup$ (red) for the $90$nm CMOS inverter chain that produced \cref{fig:wave90_inv}. 
Taken from \cite[Fig.~7]{FNNS19:TCAD}}
  \label{fig:delta90}
\end{figure}

The main reason for the dependency of the gate delay $\dup$ and $\ddo$ on the 
previous-output-to-input delay $T$, i.e., the drafting effect already covered by single-history
models like IDM and DDM, is the fact that forcing a full voltage-swing output transition to cross the threshold
voltage again takes longer than doing the same for a non-full voltage swing transition. Consider
the output signals of three stages (inv2, inv4, inv6) of an inverter chain shown in 
\cref{fig:wave90_inv}, which is taken from \cite{FNNS19:TCAD}: It is apparent that the second
pulse of inv4 is so short that its rising transition does not execute a full voltage swing
before the falling transition takes effect, which causes the latter to hit the threshold 
voltage earlier than the corresponding waveform at inv2, which is (almost) full-swing.
The consequence of this drafting effect is a decrease of $\dup$ and $\ddo$ when
$T$ becomes small, see \cref{fig:delta90}.

As already mentioned in \cref{sec:DDTA}, experiments using the Involution Tool in \cite{OMFS20:INTEGRATION} 
showed that the prediction accuracy of the simple IDM for multi-input
gates is below expectations. This is primarily due to the fact that a model of a multi-input gate that combines
single-input single-output IDM channels with zero-time Boolean gates cannot 
properly capture output delay variations caused by \emph{multiple input switching} (MIS) effects: output transitions may be sped up/slowed down when \emph{different} inputs switch in close temporal proximity~\cite{CGB01:DAC}. 

\begin{figure}[t!]
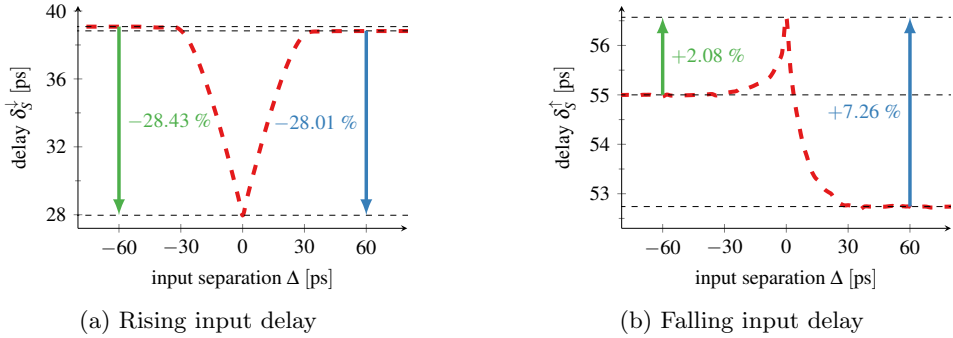

  \centering
  \subfloat[Rising input delay]{
    \includegraphics[width=0.34\linewidth]{\figPath{nor2_out_down_charlie_15nm_colored.pdf}}%
    \label{corFig3}}
  \hfil
  \subfloat[Falling input delay]{
    \includegraphics[width=0.34\linewidth]{\figPath{nor2_out_up_charlie_15nm_colored.pdf}}%
    \label{corFig5}}
  \caption{MIS effects in the measured delay of a $15$nm technology
CMOS \NOR\ gate.}\label{fig:Charlie15nmSim}
\end{figure}

Consider the CMOS implementation of a \NOR\ gate shown in \cref{fig:nor_CMOS}, which consists of two serial \pmos\ ($T_1$ and $T_2$) for charging the load capacitance $C$ (producing a rising output transition), and two parallel \nmos\ transistors ($T_3$ and $T_4$) for discharging it (producing a falling one). 
When an input experiences a rising transition, the corresponding \nmos\ transistor closes while the corresponding \pmos\ transistor opens, so $C$ will be discharged.
If both inputs $A$ and $B$ experience a rising transition at the same time, 
$C$ is discharged twice as fast. Since the gate delay depends on the 
discharging speed, it follows that the falling output delay 
$\dsd(\Delta)$ increases (by almost 30\% in the example shown in \cref{corFig3}) when the \emph{input separation time} 
$\Delta=t_B-t_A$ increases from 0 to $\infty$ or decreases from 0 to $-\infty$.
For falling input transitions, the behavior of the \NOR\ gate is quite different: \cref{corFig5} shows that the MIS effects lead to a moderate decrease of the rising output delay $\dsu(\Delta)$ when $|\Delta|$ goes from 0 to $\infty$, which is primarily caused
by capacitive coupling and the parasitic capacitances at the nodes between the \pmos\ transistors.

\section{A Thresholded Hybrid Model for Interconnected \NOR\ Gates}
\label{sec:Prel}

In this section, we summarize the cornerstones and the most important properties of the thresholded hybrid first-order ODE model for a 2-input \NOR\ gate developed in \cite{ferdowsi2024hybrid}. These results will be instrumental for developing a fully-fledged digital delay model in \cref{Sec:General_Delay_Model}. Note that it immediately allows to also
derive a digital delay model also for the ``dual'' 2-input \NAND\ gate.


\begin{figure}[h]
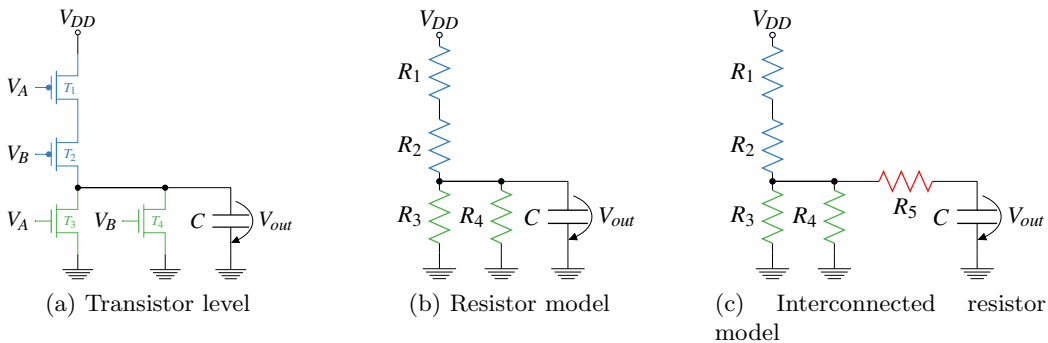

  \centering
  \subfloat[Transistor level]{
\includegraphics[height=0.23\linewidth]{\figPath{nor_RC_colored.pdf}}  
    \label{fig:nor_CMOS}}
  \hfil
  \subfloat[Resistor model]{
 \includegraphics[height=0.23\linewidth]{\figPath{nor_R_colored.pdf}}%
    \label{FigureNOR-GATE}}
  \hfil
  \subfloat[Interconnected resistor model]{
 \includegraphics[height=0.23\linewidth]{\figPath{nor_R_interconnect2.pdf}}%
    \label{FigureNOR-GATE_Int}}    
  \caption{Transistor schematic and the resistor model of a CMOS \NOR\ gate along with its augmented RC interconnect component.}
\end{figure}

The model of a 2-input CMOS \NOR\ gate proposed in \cite{ferdowsi2025faithful,ferdowsi2024hybrid} is based 
on replacing the transistors in \cref{fig:nor_CMOS} by time-varying resistors, as shown in \cref{FigureNOR-GATE}. These resistors, denoted as $R_i(t)$ for $i \in \{1,\ldots,4 \}$, vary between a predetermined on-resistance $R_i$ and the off-resistance $\infty$.
The governing principle for $R_i(t)$, which has been inspired by the Shichman-Hodges transistor model~\cite{ShichmanHodges}, is contingent upon the state of the corresponding input signal at the gate of the corresponding transistor.

This results in a hybrid model comprising four distinct modes, according to the four possible input states $(A,B)\in \{(0,0), (0,1), (1,0), (1,1) \}$. \cref{tab:T1} shows all possible input state transitions and the corresponding
resistor time evolution mode switches. Double arrows in the mode switch names
indicate MIS-relevant modes, whereas $+$ and $-$ indicate whether input $A$ switched
before $B$ or the other way round.  For instance, assume the system is in state
$(0,0)$ initially, i.e., that both $A$ and $B$ were set to 0 at time
$t_A= t_B= -\infty$. This causes $R_1$ and $R_2$ to be in the \emph{on-mode},
whereas $R_3$ and $R_4$ are in the \emph{off-mode}. If $A$ is switched to 1 at time
$t_A=0$, $R_1$ resp.\ $R_3$ is switched to the
off-mode resp.\ on-mode at time
$t^{\off}_1 = t^{\on}_3 = t_A = 0$. The corresponding mode switch is
$T_{-}^{\uparrow}$ and reaches state $(1,0)$. If $B$ is also
switched to 1 at some time $t_B=\Delta>0$, $R_2$ resp.\ $R_4$ is switched to the off-mode resp.\
on-mode at time $t^{\off}_2= t^{\on}_4 = t_B=\Delta$. The corresponding
mode switch is $T_{+}^{\uparrow\uparrow}$ and reaches state $(1,1)$. 

\begin{table}[t]
\centering
\caption{\small State transitions and modes. $\uparrow$ and $\uparrow \uparrow$ (resp.\ $\downarrow$ and $\downarrow \downarrow$) represent the first and the second rising (resp.\ falling) input transitions. $+$ and $-$ specify the sign of the switching time difference $\Delta=t_B-t_A$.}
\scalebox{0.9}
{
\begin{tabular}{lllllllllll}
\hline
Mode                            &  & Transition                &  & $t_A$       & $t_B$       &  & $R_1$                & $R_2$                & $R_3$                & $R_4$                \\ \cline{1-1} \cline{3-3} \cline{5-6} \cline{8-11} 
$T^{\uparrow}_{-}$              &  & $(0,0) \rightarrow (1,0)$ &  & $0$         & $-\infty$   &  & $on \rightarrow off$ & $on$                 & $off \rightarrow on$ & $off$                \\
$T^{\uparrow \uparrow}_{+}$     &  & $(1,0) \rightarrow (1,1)$ &  & $-|\Delta|$ & $0$         &  & $off$                & $on \rightarrow off$ & $on$                 & $off \rightarrow on$ \\
$T^{\uparrow}_{+}$              &  & $(0,0) \rightarrow (0,1)$ &  & $-\infty$   & $0$         &  & $on$                 & $on \rightarrow off$ & $off$                & $off \rightarrow on$ \\
$T^{\uparrow \uparrow}_{-}$     &  & $(0,1) \rightarrow (1,1)$ &  & $0$         & $-|\Delta|$ &  & $on \rightarrow off$ & $off$                & $off \rightarrow on$ & $on$                 \\
$T^{\downarrow}_{-}$            &  & $(1,1) \rightarrow (0,1)$ &  & $0$         & $-\infty$   &  & $off \rightarrow on$ & $off$                & $on \rightarrow off$ & $on$                 \\
$T^{\downarrow \downarrow}_{+}$ &  & $(0,1) \rightarrow (0,0)$ &  & $-|\Delta|$ & $0$         &  & $on$                 & $off \rightarrow on$ & $off$                & $on \rightarrow off$ \\
$T^{\downarrow}_{+}$            &  & $(1,1) \rightarrow (1,0)$ &  & $-\infty$   & $0$         &  & $off$                & $off \rightarrow on$ & $on$                 & $on \rightarrow off$ \\
$T^{\downarrow \downarrow}_{-}$ &  & $(1,0) \rightarrow (0,0)$ &  & $0$         & $-|\Delta|$ &  & $off \rightarrow on$ & $on$                 & $on \rightarrow off$ & $off$                \\ \hline
\end{tabular}}
\label{tab:T1}
\end{table}

For the governing principle dictating the temporal behavior of $R_i(t)$ during the on- and off-mode, 
the model employs the continuous resistance model defined by
\begin{align}
R_j^{\on}(t) &= \frac{\alpha_j}{t-t^{\on}}+R_j; \ t \geq t^{\on}, \label{on_mode}\\
R_j^{\off}(t) &= \beta_j (t-t^{\off}) +R_j; \ t \geq t^{\off}, \label{off_mode}
\end{align}
for some constant slope parameters $\alpha_j$ [\si{\ohm\s}], $\beta_j$
[\si[per-mode=symbol]{\ohm\per\s}], and on-resistance $R_j$ [\si{\ohm}] is used;
$t^{\on}$ resp.\
$t^{\off}$ represent the time when the respective transistor is switched on
resp.\ off. These equations are backed-up by the Shichman-Hodges transistor model~\cite{ShichmanHodges}, 
which assumes a quadratic correlation between the output current and the input voltage: \cref{on_mode}
and \cref{off_mode} follow from approximating the latter by 
$d \sqrt{t-t_0}$ in the operation range close to the threshold voltage $\vth$, 
with $d$ and $t_0$ denoting appropriate fitting parameters.
 
Interestingly, continuously varying resistors are only needed for switching on the \pmos\ transistors. 
All other transistor switchings, i.e., both the 
switching-off of the \pmos\ transistors
and any switching on or off of the \nmos\ transistors, happen instantaneously, 
which is accomplished by choosing the model parameters $\alpha_i=0$ in \cref{on_mode} for $i \in \{3,4 \}$,
and $\beta_k=\infty$ in \cref{off_mode} for $k \in \{1, \ldots ,4 \}$. 

Rather than including $R_1(t)$ and $R_2(t)$ in the state of the ODEs governing the 
appropriate modes of the hybrid model, which would blow-up their dimensions, 
they are incorporated by means of time-varying coefficients in simple 
first-order ODEs. In \cite{ferdowsi2024hybrid}, the resulting model has been
enhanced by also incorporating a simple lumped RC interconnect model.
\cref{FigureNOR-GATE_Int} shows the augmented schematics. 

By applying Kirchhoff's rules to \cref{FigureNOR-GATE_Int}, one derives the non-homogeneous ODE:

\begin{equation}
\label{EqIC0}
\frac{\dd V_{out}(t)}{\dd t}=-\frac{V_{out}(t)}{C\,R_g(t)(\frac{R_5}{R_g(t)}+1)}+\frac{\vdd}{C(R_1(t)+R_2(t))(\frac{R_5}{R_g(t)}+1)},
\end{equation}
where
$\frac{1}{R_g(t)}=\frac{1}{R_1(t)+R_2(t)}+\frac{1}{R_3(t)}+\frac{1}{R_4(t)}$ and
$U(t)=\frac{V_{DD}}{C(R_1(t)+R_2(t))}$. 

Although the time-dependent function $f(t)=1/(\frac{R_5}{R_g(t)}+1)$ prohibits an explicit solution of \cref{EqIC0},
it is possible to approximate $f(t)$ with mode-dependent constants. The resulting non-homogeneous non-constant coefficient
first order ODE \emph{can} be solved explicitly, leading to
\begin{equation}
\label{Eq1}
V_{out}(t)= V_0\ e^{-G(t)} + \int_{0}^{t} U(s)\ e^{G(s)-G(t)}\dd s,
\end{equation}
where $V_0=V_{out}(0)$ denotes the initial condition and $G(t) = \int_{0}^{t}
(\frac{C}{F}\,R_g(s))^{-1} \dd s$.
As comprehensively described in \cite{ferdowsi2024hybrid}, depending on each particular resistor's mode in each input state transition, different expressions for $R_g(t)$ and $U(t)$ are obtained. More specifically, each mode switch enables a specific ODE system, the solution of which provides the respective trajectory. Denoting $I_1= \int_{0}^{t} \frac{\dd s}{R_1(s)+R_2(s)}$, $I_2= \int_{0}^{t}\frac{\dd s}{R_3(s)}$, and $I_3=\int_{0}^{t} \frac{\dd s}{R_4(s)}$, \cref{T:InerInt} summarizes those. 

\begin{table}[h]
\centering
\caption{Integrals $I_1(t)$, $I_2(t)$, $I_3(t)$ and the function $U(t)$ for every possible mode switch; $\Delta=t_B-t_A$, and $2R=R_{p_A}+R_{p_B}$.}
\scalebox{0.9}
{
\begin{tabular}{llllll}
\hline
Mode                            &  & $I_1(t)= \int_{0}^{t} \frac{\dd s}{R_1(s)+R_2(s)}$                         & $I_2(t)= \int_{0}^{t}\frac{\dd s}{R_3(s)}$ & $I_3(t)=\int_{0}^{t} \frac{\dd s}{R_4(s)}$ & $U(t)= \frac{\vdd}{C(R_1(t)+R_2(t))}$                                                               \\ \cline{1-1} \cline{3-6} 
$T^{\uparrow}_{-}$              &  & $0$                                                                        & $\int_{0}^{t} (1/R_{n_A})\dd s$            & $0$                                        & $0$                                                                                                 \\
$T^{\uparrow \uparrow}_{+}$     &  & $0$                                                                        & $\int_{0}^{t} (1/R_{n_A})\dd s$            & $\int_{0}^{t} (1/R_{n_B}) \dd s$           & $0$                                                                                                 \\
$T^{\uparrow}_{+}$              &  & $0$                                                                        & $0$                                        & $\int_{0}^{t} (1/(R_{n_B}) \dd s$          & $0$                                                                                                 \\
$T^{\uparrow \uparrow}_{-}$     &  & $0$                                                                        & $\int_{0}^{t} (1/R_{n_A}) \dd s$           & $\int_{0}^{t} (1/R_{n_B}) \dd s$           & $0$                                                                                                 \\
$T^{\downarrow}_{-}$            &  & $0$                                                                        & $0$                                        & $\int_{0}^{t} (1/R_{n_B}) \dd s$           & $0$                                                                                                 \\
$T^{\downarrow \downarrow}_{+}$ &  & $\int_{0}^{t}(1/(\frac{\alpha_1}{s+\Delta}+\frac{\alpha_2}{s}+2R))\dd s$   & $0$                                        & $0$                                        & $\frac{\vdd t(t+ \Delta)}{C(2 R t^2 +(\alpha_1 + \alpha_2 + 2 \Delta R)t + \alpha_2 \Delta)}$       \\
$T^{\downarrow}_{+}$            &  & $0$                                                                        & $\int_{0}^{t} (1/(R_{n_A}) \dd s$          & $0$                                        & $0$                                                                                                 \\
$T^{\downarrow \downarrow}_{-}$ &  & $\int_{0}^{t}(1/(\frac{\alpha_1}{s}+\frac{\alpha_2}{s+|\Delta|}+2R))\dd s$ & $0$                                        & $0$                                        & $\frac{\vdd t(t+ |\Delta|)}{C(2 R t^2 +(\alpha_1 + \alpha_2 + 2 |\Delta| R)t + \alpha_1 |\Delta|)}$ \\ \hline
\end{tabular}}
\label{T:InerInt}
\end{table}

The following \cref{thm:MISOuttraj} summarizes the resulting analytic formulas for the output voltage trajectories obtained in \cite{ferdowsi2024hybrid}. It uses the notation $R_1 = R_{p_{A}}$, $R_2 = R_{p_{B}}$ with the abbreviation $2R = R_{p_{A}} + R_{p_{B}}$ for the two \pmos\ transistors $T_1$ and $T_2$ and $R_3 = R_{n_{A}}$, $R_4 = R_{n_{B}}$ for the two \nmos\ transistors $T_3$ and $T_4$. 

\begin{prp}[Output voltage trajectories for the interconnect-augmented \NOR\ gate {\cite[Thm.~1]{ferdowsi2024hybrid}}] \label{thm:MISOuttraj}
For any $0 \leq |\Delta| \leq \infty$, the output voltage trajectory functions of our model for rising input transitions are given by

\begin{flalign}
V_{out}^{T^{\uparrow}_{-}}(t) &= V_{out}^{T^{\uparrow}_{-}}(0) e^{\frac{-t}{C_1 R_{n_{A}}}},
\label{outsig1}\\
V_{out}^{T^{\uparrow}_{+}} (t) &= V_{out}^{T^{\uparrow}_{+}}(0) e^{\frac{-t}{C^{'}_{1} R_{n_{B}}}},
\label{outsig1neg}\\
V_{out}^{T^{\uparrow \uparrow}_{+}}(t) &=V_{out}^{T^{\uparrow}_{-}} (\Delta)  e^{- \bigl(\frac{1}{C_2R_{n_A}}+\frac{1}{C_2R_{n_B}}\bigr)t},
\label{outsig2}\\
V_{out}^{T^{\uparrow \uparrow}_{-}}(t) &=V_{out}^{T^{\uparrow}_{+}} (\Delta)  e^{- \bigl(\frac{1}{C_2R_{n_A}}+\frac{1}{C_2R_{n_B}}\bigr)t}.
\label{outsig2_neg}
\end{flalign}

The output voltage trajectory functions for falling input transitions are given by

\begin{flalign}
V_{out}^{T^{\downarrow}_{-}}(t) &= V_{out}^{T_{-}^{\downarrow}}(0) e^{\frac{-t}{C^{'}_{1}R_{n_B}}},
\label{eq:FirstFalltheorem}\\
V_{out}^{T^{\downarrow}_{+}}(t) &= V_{out}^{T_{+}^{\downarrow}}(0) e^{\frac{-t}{C_{1}R_{n_A}}},
\label{eq:FirstFalltheorem_plus}\\
V_{out}^{T^{\downarrow \downarrow}_{+}}(t)&= \vdd  + \bigl(V_{out}^{T^{\downarrow}_{-}}(\Delta) -\vdd   \bigr) \cdot\left[ e^{\frac{-t}{2RC_3}} \Bigl(1+\frac{2t}{d+\sqrt{\chi}}\Bigr)^{\frac{-A+a}{2RC_3}} \Bigl(1+\frac{2t}{d-\sqrt{\chi}}\Bigr)^{\frac{A}{2RC_3}} \right],  \label{SoughtOutput}
\end{flalign}

where $a=\frac{\alpha_1+\alpha_2}{2R}$, $d=a+\Delta$,  $ \chi=d^2-4c'$, $c'=\frac{\alpha_2 \Delta}{2R}$, and $A=\frac{\alpha_2\Delta - aR(d- \sqrt{\chi})}{2R\sqrt{\chi}}$. 
Besides,

\begin{align}
C_1&=\frac{C(R_5+R_{n_{A}})}{R_{n_{A}}}, \label{eq:C_1_int}\\
C_1^{'} &=\frac{C(R_5+R_{n_{B}})}{R_{n_{B}}}, \label{eq:C_1_prime_int}\\
C_2 &=\frac{C\bigl(R_5(R_{n_{A}}+R_{n_{B}})+R_{n_{A}}R_{n_{B}}\bigl)}{R_{n_{A}}R_{n_{B}}}, \label{eq:C_2_int}\\
C_3&=\frac{C(R_5+2R)}{2R}. \label{eq:C_3_int}
\end{align}

The output voltage trajectory $V_{out}^{T^{\downarrow \downarrow}_{-}}(t)$ for
negative $\Delta$ is
obtained from \cref{SoughtOutput} by exchanging $\alpha_1$ and $\alpha_2$, replacing
$V_{out}^{T^{\downarrow}_{-}}(\Delta)$ by $V_{out}^{T^{\downarrow}_{+}}(|\Delta|)$, and $\Delta$ by $|\Delta|$ in $d$, $\chi$ and $A$. Defining $\bd=a+|\Delta|$,  $ \bchi=\bd^2-4\bc'$, $\bc'=\frac{\alpha_1 |\Delta|}{2R}$, and $\bA=\frac{\alpha_1|\Delta| - aR(\bd- \sqrt{\bchi})}{2R\sqrt{\bchi}}$ accordingly, we obtain

\begin{flalign}
V_{out}^{T^{\downarrow \downarrow}_{-}}(t)&= \vdd  + \bigl(V_{out}^{T^{\downarrow}_{+}}(|\Delta|) -\vdd   \bigr) \cdot\left[ e^{\frac{-t}{2RC_3}} \Bigl(1+\frac{2t}{\bd+\sqrt{\bchi}}\Bigr)^{\frac{-\bA+a}{2RC_3}} \Bigl(1+\frac{2t}{\bd-\sqrt{\bchi}}\Bigr)^{\frac{\bA}{2RC_3}} \right]. \label{SoughtOutputneg}
\end{flalign}

\end{prp}

The trajectories $V_{out}^{T^{\uparrow \uparrow}_{+}}(t)$, $V_{out}^{T^{\uparrow \uparrow}_{-}}(t)$
and $V_{out}^{T^{\downarrow \downarrow}_{+}}(t)$, $V_{out}^{T^{\downarrow \downarrow}_{-}}(t)$ in
\cref{thm:MISOuttraj} are specifically tailored to facilitate the computation of
the MIS delays.
More specifically, determining the delay expressions requires inverting these trajectories. For example, for $\Delta \geq 0$, the delay can be obtained according to the following two procedures:
\begin{itemize}
\item[(i)] Compute $V_{out}^{T^{\uparrow}_{-}} (\Delta)$ for the first transition $(0,0)\to(1,0)$
  and use it as the initial value for $V_{out}^{T^{\uparrow \uparrow}_{+}}(t)$ governing the second
  transition $(1,0)\to(1,1)$; the ultimately sought MIS delay $\delta_{M,+}^{\downarrow}(\Delta)$ is
  the time (measured from the first transition until either the first or the second trajectory
  crosses the threshold voltage $\vdd/2$ from
  above when the first one starts from $V_{out}^{T^{\uparrow}_{-}} (0)=\vdd$.
\item[(ii)] Compute $V_{out}^{T^{\downarrow}_{-}}(\Delta)$ for the first transition $(1,1)\to(0,1)$
  and use it as the initial value for
  $V_{out}^{T^{\downarrow \downarrow}_{+}}(t)$ governing the second transition $(0,1)\to(0,0)$;
  the ultimately sought MIS delay $\delta_{M,+}^{\uparrow}(\Delta)$
  is the time (measured from the second transition) until the second trajectory
  crosses the threshold voltage $\vdd/2$ from below when the first one starts from
  $V_{out}^{T^{\downarrow}_{-}}(0)=0$.
\end{itemize}

Clearly, to obtain the MIS delays, it suffices to stabilize the system at $(1,1)$ resp. $(0,0)$ for the case of rising resp. falling input transitions, resulting from the assumption $T \rightarrow \infty$. This equivalently leads to having $V_{out}^{T^{\uparrow}_{-}}(0)= V_{out}^{T^{\uparrow}_{+}}(0)=\vdd$ and $V_{out}^{T_{-}^{\downarrow}}(0)=V_{out}^{T_{+}^{\downarrow}}(0)=0$. Although, with this in mind, inverting the appropriate trajectory formula was straightforward for the rising input transition case (i), it turned out to be challenging for the falling input transition case (ii). Whereas  a complex piecewise approximation (in terms of $\Delta$) was used for both the trajectory and the corresponding delay formula was used in \cite{ferdowsi2023accurate}, explicit trajectory formulas developed in \cite{ferdowsi2024hybrid}, stated in \cref{thm:MISOuttraj}, allowed the derivation of explicit MIS delay formulas. \cref{Corol:1} summarizes these formulas.

\begin{prp}[MIS delay functions for the interconnect-augmented \NOR\ gate {\cite[Thm.~2]{ferdowsi2024hybrid}}]\label{Corol:1}
For any $0 \leq |\Delta| \leq \infty$, the MIS delay functions, for rising and falling input transitions, 
of our interconnect-augmented model with pure delay $\dmin\geq 0$ are respectively given by

\begin{align}
\delta_{M,+}^{\downarrow}(\Delta) = \begin {cases}
 \frac{\log(2)C_2R_{n_A}R_{n_B} - \frac{C_2}{C_1} \Delta R_{n_B}}{R_{n_A}+R_{n_B}} + \Delta + \dmin &   \ \ 0 \leq \Delta < \log(2)C_1R_{n_A}  \\ 
 \log(2)C_1R_{n_A} + \dmin &   \ \ \Delta \geq \log(2)C_1R_{n_A}
\end {cases} \label{Fallingmisdelayformula_int}
\\
\delta_{M,-}^{\downarrow}(\Delta) = \begin{cases}
 \frac{\log(2)C_2R_{n_A}R_{n_B} + \frac{C_2}{C_1'} |\Delta| R_{n_A}}{R_{n_A}+R_{n_B}} + |\Delta| + \dmin &   \ \ |\Delta| < \log(2)C_1^{'}R_{n_B}  \\ 
\log(2)C_1^{'}R_{n_B} + \dmin &   \ \ |\Delta| \geq \log(2)C_1^{'}R_{n_B}
\end {cases} \label{Fallingmisdelayformulaminus_int}
\end{align}

\begin{align}
\delta_{M,+}^{\uparrow}(\Delta) &= \begin {cases}
\delta_{0} - \frac{\alpha_1}{\alpha_1+\alpha_2} \Delta + \dmin  &   \ \ 0 \leq \Delta < \frac{(\alpha_1+\alpha_2)(\delta_{0} - \delta_{\infty})}{\alpha_1}   \\ 
\delta_{\infty} + \dmin &   \ \ \Delta \geq \frac{(\alpha_1+\alpha_2)(\delta_{0} - \delta_{\infty})}{\alpha_1}
\end {cases}\label{Risingmisdelayformula_int}
\\
\delta_{M,-}^{\uparrow}(\Delta) &= \begin {cases}
\delta_{0} - \frac{\alpha_2}{\alpha_1+\alpha_2} |\Delta| + \dmin &   \ \ 0 \leq |\Delta| < \frac{(\alpha_1+\alpha_2)(\delta_{0} - \delta_{-\infty})}{\alpha_2}  \\ 
\delta_{-\infty}' + \dmin&   \ \ |\Delta| \geq \frac{(\alpha_1+\alpha_2)(\delta_{0} - \delta_{-\infty})}{\alpha_2}
\end {cases}\label{Risingmisdelayformulaminus_int}
\end{align}
where $C_1$, $C_1'$, $C_2$, and $C_3$ are given by \cref{eq:C_1_int}, \cref{eq:C_1_prime_int},
\cref{eq:C_2_int}, and \cref{eq:C_1_int}, respectively, and

\begin{align}
&\delta_{0}= - \frac{\alpha_1 + \alpha_2}{2R} \biggl[ 1+ W_{-1}\biggl(\frac{-1}{e \cdot 2^{\frac{4R^2C_3}{\alpha_1+ \alpha_2}}}\biggr) \biggr],  \label{eq:delta0_int} \\
&\delta_{\infty}= -\frac{\alpha_2}{2R} \biggl[ 1+ W_{-1}\biggl(\frac{-1}{e \cdot 2^{\frac{4R^2C_3}{\alpha_2}}}\biggr) \biggr],  \label{eq:deltainf_int} \\
&\delta_{-\infty}= -\frac{\alpha_1}{2R} \biggl[ 1+ W_{-1}\biggl(\frac{-1}{e \cdot 2^{\frac{4R^2C_3}{\alpha_1}}}\biggr) \biggr]. \label{eq:deltaminf_int}
\end{align}
\end{prp}

A non-zero pure delay term $\dmin>0$,
already foreseen in the original IDM model \cite{FMNNS18:DATE}, is essential for
guaranteeing causality, and may also help for model parametrization. Obviously, for the delay formulas expressed in 
\cref{Corol:1} to be useful, it is mandatory to have a practical procedure for determining all the model parameters.
Given the MIS delay values of some real interconnected \NOR\ gate, namely, $\ddoD_S(-\infty)$, $\ddoD_S(0)$, and $\ddoD_S(\infty)$ 
according to \cref{corFig3} and $\dupD_S(-\infty)$, $\dupD_S(0)$, and $\dupD_S(\infty)$ according to \cref{corFig5},
it is desired to determine suitable values for the parameters $\alpha_1$, $\alpha_2$, $C$, $R$, $R_{n_A}$, $R_{n_B}$, and $R_5$
such that the MIS delays predicted by our model \emph{match} these values, in the sense that 
$\ddoD_{M,-}(-\infty)=\ddoD_S(-\infty)$, $\ddoD_{M,-}(0)=\ddoD_{M,+}(0)=\ddoD_S(0)$, $\ddoD_{M,+}(\infty)=\ddoD_S(\infty)$ and $\dupD_{M,-}(-\infty)=\dupD_S(-\infty)$,  $\dupD_{M,-}(0)=\dupD_{M,+}(0)=\dupD_S(0)$, $\dupD_{M,+}(\infty)=\dupD_S(\infty)$. 
The modeling in \cite{ferdowsi2024hybrid} assumes that the pure delay $\dmin>0$ is 
part of the interconnect at the gate output, and is just subtracted from the given MIS delay
values before the matching. When using the model, $\dmin$ must hence be added to the calculated delay values. This is why an additional $\dmin$ appears in all delay expressions in \cref{Corol:1} as well as in all subsequent delay formulas. Note that its actual value has been found to be uncritical w.r.t.\ successful parametrization, however. 

\cref{thm:gatechar} summarizes the procedure developed in \cite{ferdowsi2024hybrid} for parametrizing the introduced delay model for interconnected \NOR\ gates.

\begin{prp}[Model parametrization for interconnect-augmented \NOR\ gates {\cite[Thm.~3]{ferdowsi2024hybrid}}]\label{thm:gatechar}
Let $\dmin\geq 0$ be some additional interconnect pure delay and $\ddoD_S(-\infty)$,  $\ddoD_S(0)$, $\ddoD_S(\infty)$ and $\dupD_S(-\infty)$,  $\dupD_S(0)$, $\dupD_S(\infty)$ be the MIS delay values of a real interconnected \NOR\ gate that shall be matched by our model. Given an arbitrary value $C$ for the load capacitance, this is accomplished by choosing the model parameters as follows:

\begin{flalign}
&R_5 =  \frac{\bigl( \ddoD_S(0) - \dmin - \epsilon \bigr)}{\log(2)C}   \label{eq:dmin}&\\
&R_{n_{A}}=\frac{\ddoD_S(\infty)-\ddoD_S(0)+\epsilon}{\log(2) C} \label{eq:rna}&\\
&R_{n_{B}}=\frac{\ddoD_S(-\infty)-\ddoD_S(0)+\epsilon}{ \log(2) C} \label{eq:rnb}&  \\
&\epsilon=\sqrt{\bigl(\ddoD_S(\infty)-\ddoD_S(0)\bigr)\bigl(\ddoD_S(-\infty)-\ddoD_S(0)\bigr)}
\end{flalign}

Furthermore, using the function
\begin{flalign}
&A(t,R,R_5,C)=& \nonumber \\
&\frac{-2R \bigl(t-C(R_5+2R) \cdot \log(2) \bigr)}{W_{-1}\Bigl( \bigl(\frac{C(R_5+2R) \cdot \log(2)}{t}-1\bigr) e^{\frac{C(R_5+2R) \cdot \log(2)}{t}-1} \Bigl) +1 - \frac{C(R_5+2R) \cdot \log(2)}{t}}\label{eq:AtRC},&
\end{flalign}

determine $R$ by numerically\footnote{Whereas there might be a way to solve it analytically, we did not find it so far.} solving the equation 
\begin{flalign}
&A\bigl(\dupD_S(0)-\dmin,R,R_5,C\bigr)-A\bigl(\dupD_S(\infty)-\dmin,R,C\bigr)&\nonumber \\
& - A\bigl(\dupD_S(-\infty)-\dmin,R,C\bigr) = 0 \label{eq:forR}, &
\end{flalign}
and finally choose
\begin{flalign}
\alpha_1 &= A\bigl(\dupD_S(-\infty)-\dmin,R,R_5,C\bigr)\label{eq:alpha1},\\
\alpha_2 &= A\bigl(\dupD_S(\infty)-\dmin,R,R_5,C\bigr)\label{eq:alpha2}.
\end{flalign}
\end{prp}

We conclude this section by stressing that our focus on simple first-order
ODE modes is a deliberate one. We are of course aware of the fact that
such models cannot cover higher-order effects, like e.g.\
the Miller effect in transistors, capacitive coupling, and distributed
RC and even inductances arising in real interconnects. Similarly, our
focus on thresholded hybrid models rules does not allow us to cover
slew-related effects. The main reason for those restrictions is our
goal to develop analytic closed-form delay expressions, which requires
quite some effort
already for our first-order models, cf.\ \cite{ferdowsi2024faithful}, and 
seems virtually impossible for higher-order models.

At the same time, we are of course very interested in the accuracy
of the resulting model predictions compared to real circuits,
where all the above effects do of course occur. An important part of
our research is hence devoted to the experimental accuracy validation, like
the one of \cref{Sec:experiments}. Overall, all our experimental
results obtained so far revealed a surprisingly good accuracy,
which suggests that higher-order effects are not dominant in ``typical''
settings. 


\section{Full Delay Model for \NOR\ Gates}
\label{Sec:General_Delay_Model}


In this section, we will develop a full delay model for (interconnected) \NOR\ gates, i.e.,
one that incorporates both drafting and MIS effects: Unlike the delay formulas provided
in \cref{Corol:1}, which depend on $\Delta$ only, the ones stated in \cref{Theorem:Del_Traj_NOR}
will depend on both $\Delta$ and $T$. In the case of the IDM \cite{FNNS19:TCAD}, a single
$\dup(T)$ and $\ddo(T)$ suffices, with $T$ respresenting the \emph{previous-output-to-input 
delay}, i.e., $T=t-t_{o}'$, where $t$ is the time of the current input transition and $t_o'$ is the
time of the output transition corresponding to the previous input transition. 
In the case of a \NOR\ gate, however, multiple expressions for both $\dup(\Delta,T)$ 
and $\ddo(\Delta,T)$ are needed, depending on the previous input transitions, with
possibly varying definitions of the parameters $\Delta$ and $T$.

More specifically, for just dealing with MIS effects, only four types of transitions $(0,0) \rightarrow (1,0) \rightarrow (1,1)$, $(0,0) \rightarrow (0,1) \rightarrow (1,1)$, $(1,1) \rightarrow (0,1) \rightarrow (0,0)$, and $(1,1) \rightarrow (1,0) \rightarrow (0,0)$ need to be considered. Moreover, it suffices to consider the situation where the first mode (i.e., $(0,0)$ and $(1,1)$ for the rising resp.\ falling input transition cases) is already assumed at time $-\infty$, which entirely rules out the drafting effect. As a consequence, $T=\infty$ here, and the second transition starts from the trivial initial voltage value $V_{out}^{T^{\uparrow}_{-}}(0)=V_{out}^{T^{\uparrow}_{+}}(0)=\vdd$ resp.\ $V_{out}^{T^{\downarrow_{-}}}(0)=V_{out}^{T^{\downarrow}_{+}}(0)=0$.

Also incorporating the drafting effect implies that the time the first mode is assumed is not $-\infty$, but could be anything, which leads to $T<\infty$ and arbitrary initial voltage values. In order to compute the respective gate delays $\dup(\Delta,T)$ and $\ddo(\Delta,T)$
(by inverting the trajectories given in \cref{thm:MISOuttraj}), these initial values need to be determined analytically. Equipped with this information (see \cref{Theorem:Del_Traj_NOR} below), given an arbitrary sequence of input transitions, the corresponding sequence of output transitions of a \NOR\ gate can be computed by iteratively applying the appropriate delay formula.

\begin{figure}[t!]
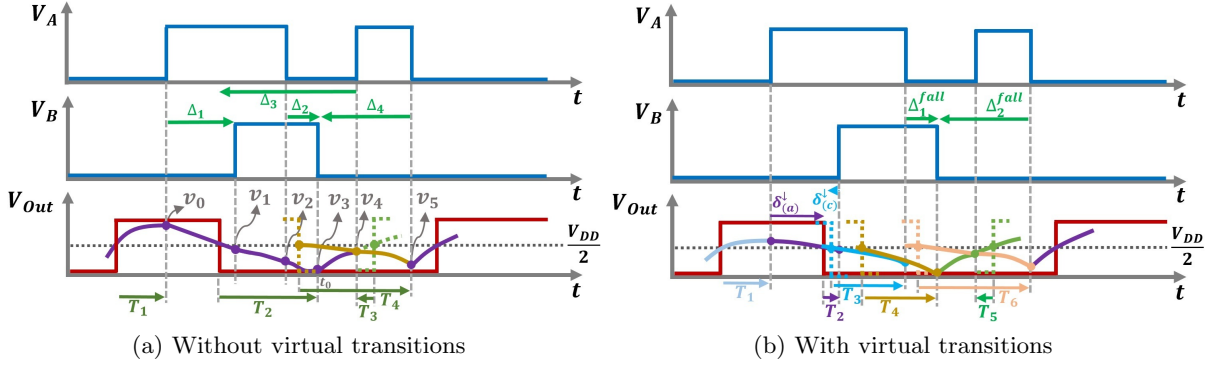

  \centering
  \subfloat[Without virtual transitions]{
    \includegraphics[width=0.49\linewidth]{\figPath{Trouble_Pulse2.pdf}}%
    \label{fig:TroubleSig}}
  \hfil
  \subfloat[With virtual transitions]{
    \includegraphics[width=0.49\linewidth]{\figPath{Trouble_Pulse4.pdf}}%
    \label{fig:TroubleSig2}}
  \caption{
(a) An example \NOR\ gate timing diagram along with some relevant $\Delta$ and $T$ parameters values, assuming $\dmin=0$ for simplicity. The purple curve specifies the analog voltage trajectory. Note that $\dmin > 0$ would cause a constant right-shift of the digital (but not the analog) output signal. Computing the initial voltage $v_3$ requires to know $v_2, v_1$ and $v_0$. The dashed green transition shows an planned output transition caused by the second falling input transition occurring at time $t_0$, which is canceled by the virtual dashed brown falling output transition caused by the second rising transition of input $A$.  (b) The sequence of output transitions along with the relevant $T$ and $\Delta$ parameters
for the sequence of input transitions in \cref{fig:TroubleSig}, in the generalized model where any transition on 
any input leads to either a real (solid) or virtual (dashed) output transition. Note the negative signs for $\Delta^{fall}_{2}$ and $T_5$. For the latter, the second rising input transition at input $A$ occurs before the previous (virtual) output transition (dotted green transition), making $T_5$ negative. A $\Delta$-dependency is only present for falling input transitions, which we hence denote by $\Delta^{fall}$.
  }\label{fig:TroubleSigMain}
\end{figure}

The main technical difficulty in developing these delay formulas is the computation of the initial voltage value
for a given transition, i.e., $V_{out}^{T^{\uparrow}_{-}}(0)$ and $V_{out}^{T^{\uparrow}_{+}}(0)$ in \cref{outsig1} and \cref{outsig1neg} and $V_{out}^{T^{\downarrow_{-}}}(0)$ and $V_{out}^{T^{\downarrow}_{+}}(0)$ in \cref{eq:FirstFalltheorem} and \cref{eq:FirstFalltheorem_plus}, depending on the previous transition(s) and the relevant parameters $T$ and $\Delta$. This becomes particularly troublesome if the parameter $T$ refers to a previous ``real'' output transition, 
i.e., the one where the output changed its state. In this case, any number of input transitions may occur in-between the
input transition that causes the output to change its state for the $n$-th time and the input transition that caused
the $(n-1)$-st state change of the output, which do not cause the output to change its state but only change the slope 
of the trajectory. \cref{fig:TroubleSig} illustrates both such a scenario (1)  and  a possible cancellation (2) of a ``planned'' output transition. In more detail:
\begin{enumerate}
\item[(1)] For computing the delay of the second falling input transition at time $t_0$, one not
only needs $\Delta_2$ and $T_2$, but also the initial voltage $v_3$. Computing the latter
not only requires the knowledge of additional input timing parameters, but also $v_2$, which 
in turn depends on the initial value $v_1$ (which in turn depends on $v_0$). Consequently, 
we cannot hope to get a delay formula that depends only on $\Delta_2$ and $T_2$.
\item [(2)] The short duration between $t_0$ and the third rising input transition (second rising transition at input $A$) does not allow the analog output signal to cross $\vdd/2$, i.e., $v_4< \vdd/2$, resulting in a \emph{cancellation}: viewed at the level of digital signals, the planned rising output transition (the green dashed transition) does not occur, but is rather
canceled by a \emph{virtual} falling output transition (the brown dashed transition) caused by the second rising transition of $A$ that is scheduled earlier. 
\end{enumerate}

To circumvent the problem of going back arbitrarily in the input transition sequence in order to compute
some initial value, we just generalize the virtual transitions already used for representing cancellations:
We just assume that \emph{any} input transition causes a corresponding output transition, albeit not necessarily
alternating ones. For example, we let the first rising transition of input $B$ also cause an
output transition, a falling one, the occurrence time of which can be
determined by continuing the output trajectory back in time until
it hits $\vdd/2$, see the light blue trajectory in \cref{fig:TroubleSig2} for an illustration. Since the (earlier) 
first rising transition of input $A$ already caused a falling output transition,
this leads to two consecutive falling transitions. This does not matter, though, since only the earliest
one causes a real output transition, whereas the later one(s) are virtual only.

As a consequence of the resulting 1:1 relation between the input transitions and the output transitions, 
it is possible to define a meaningful delay for \emph{every} input transition, as the time
difference between the input transition and the time the output voltage will cross (or has already
crossed) $\vdd/2$. Moreover, the input-to-previous-output parameter $T$ is well-defined again. 
What still needs
to be differentiated, though, is the type of the current input transition (rising, falling) and the type
of the previous input transitions of both inputs. In the case of a two-input \NOR\ gate, 
this results in eight different scenarios, which we denote by \emph{Case~(a)} to \emph{Case~(h)}
later on, listed in \cref{table:case}.

\begin{table}[htb]
\centering
\caption{\small\em Different eight cases associated with different input transitions used in \cref{Alg:NOR}.}
\label{table:case}
\scalebox{1}{
\begin{tabular}{c|c|c}
Case & Notation            & State transition            \\ \hline \hline
(a)     & $(\uparrow,0-,t)$   & $(0,0) \rightarrow (1,0)$   \\
(b)     & $(0-,\uparrow,t)$   & $(0,0) \rightarrow (0,1)$   \\
(c)     & $(1-,\uparrow,t)$   & $(1,0) \rightarrow (1,1)$   \\
(d)     & $(\uparrow,1-,t)$   & $ (0,1) \rightarrow (1,1)$  \\
(e)     & $(\downarrow,1-,t)$ & $ (1,1) \rightarrow (0,1) $ \\
(f)     & $(1-,\downarrow,t)$ & $ (1,1) \rightarrow (1,0)$  \\
(g)     & $(0-,\downarrow,t)$ & $ (0,1) \rightarrow (0,0)$  \\
(h)     & $(\downarrow,0-,t)$ & $ (1,0) \rightarrow (0,0)$  \\ \hline
\end{tabular}}
\end{table}

Whereas the parameter $T$ is the same in every case, the definition of $\Delta$ may
be different. To be more specific, in the case of rising input transitions, $\Delta$ 
no longer plays any role, i.e., $\ddo(\Delta,T)=\ddo(T)$ is independent of $\Delta$. 
Although this may seem counterintuitive, the dependency is implicit and embedded in $T$. 
The underlying reason is that a rising transition at either input causes a falling
exponential voltage trajectory, see \cref{outsig1}--\cref{outsig2_neg}, which composes
nicely. This is not the case for falling input transitions, however,
where only the second transition causes the output to actually rise. The
complex trajectory \cref{SoughtOutput} prohibits a simple composition and thus
leads to a non-trivial dependency on $\Delta$.

Another important consequence of the 1:1 relation between the input transitions and the output 
transitions is that it facilitates a time shift that greatly simplifies computing the sought initial 
voltage values. More precisely, rather than using trajectories that 
start from arbitrary initial values at time 0 (as given in \cref{thm:MISOuttraj}), we can restrict 
our attention to trajectories that virtually start from the fixed threshold voltage $\vdd/2$, and
just trace them up to time $t=\dmin+T$ for determining the sought initial voltage value at the current
input transition time. Note carefully that we do, of course, not modify the actual starting times 
of the trajectories, but rather take the (uniquely determined) time where they hit $\vdd/2$ as virtual starting times. This shifting trick will allow us to express the sought initial values as a function in terms of $T$ (and, in the case of falling input transitions, of $\Delta$).

Finally, we note that pulse cancellation in multi-input gates like our \NOR\ gate is also slightly richer than in the simple IDM. In the latter, an output transition cancellation occurs if two successive input transitions are too close, which according to $\delta(T)$, causes a re-ordering of the corresponding output transitions. In our hybrid delay model for the \NOR\ gate, a cancellation can also result from a transition of the other input. The green trajectory in \cref{fig:TroubleSig2} exemplifies this type of cancellation: the quite early second rising transition of input $A$ does not provide enough time for the output trajectory to reach $\vdd/2$, leading to a cancellation of the scheduled dotted green rising output transition.

\subsection{Simulation algorithm}

\begin{algorithm}[h!]
\tiny
\SetKwInput{Input}{Input}
\SetKwInput{Output}{Output}
\Input{A sequence of input transitions $I= ((x_i,y_i,t_i))_{i \in [0,N]}$, $N\geq 2$.}
\Output{A sequence of real output transitions $O=((o_i,t_i'))_{i \in [1,N']}$.}
$T \leftarrow  \infty$; \ $\Delta \leftarrow \infty$; \ $O \leftarrow \emptyset$; \ $\Delta_{temp}^{e/f} \leftarrow - \infty$ \\
$(x,y, t) \leftarrow (x_0,y_0, t_0)$ ; // Initial state, at $t_0=-\infty$\\
\If{$(x = 0-) \  \land \ (y = 0-)$}
{
add $(1,-\infty)$ to $O$; \ $V_{int} \leftarrow \vdd$;
}
\Else
{
add $(0,-\infty)$ to $O$; \ $V_{int} \leftarrow 0$;
}
$(x,y, t^{\text{next}}) \leftarrow get\_next\_input()$;\\
\While {$t^{\text{next}} < t_{N}$}
{
Case: $(a)=(\uparrow, 0-,t^{\text{next}})$, $(b)=(0-, \uparrow,t^{\text{next}})$, $(c)=(1-, \uparrow,t^{\text{next}})$, $(d)=(\uparrow, 1-,t^{\text{next}})$, $(e)=(\downarrow, 1-,t^{\text{next}})$, $(f)=(1-,\downarrow,t^{\text{next}})$, $(g)= (0-, \downarrow,t^{\text{next}})$, $(h)=(\downarrow, 0-,t^{\text{next}})$; \label{line:Allcases}\\
\If{Case(i) for $i \in \{ a,b,c,d\}$}
{
$t^{\text{current}} \leftarrow t^{\text{next}}$;\\ 
$t^{o} \leftarrow t^{\text{current}}+ \delta^{\downarrow}_{(i)}(V_{int})$;\label{line:computeto}\\
add $(0,t^{o})$ to $O$;\\
$(x,y, t^{\text{next}}) \leftarrow get\_next\_input()$;\\
\If{$t^{o} -\dmin < t^{\text{current}} \lor t^{o} -\dmin > t^{\text{next}}$}
{
remove $(0,t^{o})$ from $O$;\label{line:cancel1}\\
}
$T \leftarrow t^{\text{next}}- t^{o}$;\\
$V_{int} \leftarrow V_{out}^{(i)}(T+\dmin)$;\label{line:computeVinta}\\
}
\ElseIf{Case(i)  for $i \in \{ e, f \}$}
{
$t^{\text{current}} \leftarrow t^{\text{next}}$;\\ 
$\Delta^{i}_{temp} \leftarrow t^{\text{current}}$; // For computing $\Delta$ in Case~(g) or Case~(h)\\ 
$t^{o} \leftarrow t^{\text{current}}+ \delta^{\downarrow}_{(i)}(V_{int})$; \\
add $(0,t^{o})$ to $O$;\\
$(x,y, t^{\text{next}}) \leftarrow get\_next\_input()$;\\
\If{$t^{o} -\dmin< t^{\text{current}} \lor t^{o} -\dmin > t^{\text{next}}$}
{
remove $(0,t^{o})$ from $O$;\label{line:cancel2}\\
}
$T \leftarrow t^{\text{next}}- t^{o}$;\\
$V_{int} \leftarrow V_{out}^{(i)}(T+\dmin)$;\label{line:computeVinti}\\
}
\ElseIf{Case(g)}
{
$t^{\text{current}} \leftarrow t^{\text{next}}$;\\ 
$\Delta \leftarrow t^{\text{current}}- \Delta^{e}_{temp}$;\\
$\Delta^{f}_{temp} \leftarrow t^{\text{current}}$; // For computing $\Delta$ in Case~(h)\\
$t^{o} \leftarrow t^{\text{current}}+ \delta^{\uparrow}_{(g)}(\Delta,V_{int})$;\\
add $(0,t^{o})$ to $O$;\\
$(x,y, t^{\text{next}}) \leftarrow get\_next\_input()$;\\
\If{$t^{o} -\dmin < t^{\text{current}} \lor t^{o} -\dmin > t^{\text{next}}$}
{
remove $(1,t^{o})$ from $O$;\label{line:cancel3}\\
}
$T \leftarrow t^{\text{next}}- t^{o}$;\\
\If{$V_{int} \leq \vdd/2$}
{
$V_{int} \leftarrow V_{out}^{(g)}(T+\dmin,\Delta, V_{int})$;\\
}
\Else{
$V_{int} \leftarrow V_{out}^{(g)}\Bigl(T-2RC_3 \log \bigl(\vdd /(2(\vdd-V_{int})) \bigr)+\dmin,\Delta, V_{int}\Bigr)$; \label{line:caseg}\\
}
}
\ElseIf{Case(h)}
{
$t^{\text{current}} \leftarrow t^{\text{next}}$;\\ 
$\Delta \leftarrow  \Delta^{f}_{temp} -  t^{\text{current}}$;\\
$\Delta^{e}_{temp} \leftarrow t^{\text{current}}$; // For computing $\Delta$ in Case~(g)\\
$t^{o} \leftarrow t^{\text{current}}+ \delta^{\uparrow}_{(h)}(|\Delta|,V_{int})$;\\
add $(0,t^{o})$ to $O$;\\
$(x,y, t^{\text{next}}) \leftarrow get\_next\_input()$;\\
\If{$t^{o} -\dmin < t^{\text{current}} \lor t^{o} -\dmin> t^{\text{next}}$}
{
remove $(1,t^{o})$ from $O$;\label{line:cancel4}\\
}
$T \leftarrow t^{\text{next}}- t^{o}$;\\

\If{$V_{int} \leq \vdd/2$}
{
$V_{int} \leftarrow V_{out}^{(h)}(T+\dmin, \Delta,V_{int})$;\\
}
\Else
{
$V_{int} \leftarrow V_{out}^{(h)}\Bigl(T-2RC_3 \log \bigl(\vdd /(2(\vdd-V_{int})) \bigr)+\dmin,\Delta, V_{int}\Bigr)$;\\
}
}
} 
\textbf{Return} O;\\
\caption{ \scriptsize Computing the output signal trace of a \NOR\ gate.}
\label{Alg:NOR}
\end{algorithm}

We implemented the approach described above in \cref{Alg:NOR}, which allows to compute the sequence of all real output transitions $O=((o_i,t_i'))_{i \in [0,N']}$, given the sequence of the input's signal transitions $I= ((x_i,y_i,t_i))_{i \in [0,N]}$. The number
of transitions $N\geq 2$ can be finite or infinite, $N' \leq N$ is the number of real output transitions $o_i \in \{0,1 \}$
generated, and $t_0'=t_0=-\infty$. The inputs are chosen from $x_i,y_i \in \{ \uparrow, \downarrow, 0-, 1- \}$, with the restriction that $(x_0,y_0,t_0)$ with $x_0,y_0 \in \{0-, 1-\}$ gives the initial input state at time $t_0=-\infty$, whereas for $i \in [1,N]$, either input $x_i$ or input $y_i$ must be picked 
from $\{ \uparrow, \downarrow\}$ (indicating a rising or falling input at time $t_i$), whereas the other 
input must be taken from $\{0-, 1-\}$ (indicating that the input is and remains at $0$ resp.\ $1$ at time $t_i$). 
For instance, Case~(a) $(\uparrow, 0-,t_i)$ in \cref{table:case} represents the transition $(0,0) 
\rightarrow (1,0)$ occurring at time $t_i$.

Informally, \cref{Alg:NOR} works as follows: After determining the initial output state $(o_0,-\infty)$ and the
initial voltage value $V_{int}$ from the initial input state $(x_0,y_0,-\infty)$, the function $get\_next\_input()$ is used to read one input
transition after the other. Depending on the case (determined according to \cref{line:Allcases}, which
just encodes \cref{table:case}) implied by the current input transition at $t^{\text{current}}$, 
the appropriate parameters $T$ and $\Delta$ are determined. Then,
the delay formulas supplied by \cref{Theorem:Del_Traj_NOR} are used to determine the next scheduled
output transition time $t^o$. Depending on the next input transition time $t^{\text{next}}$, obtained 
via the next call of $get\_next\_input()$, a real output transition may or may not be added to $O$, depending on whether a cancellation happens or not.
Finally, the trajectory formulas also supplied by \cref{Theorem:Del_Traj_NOR} are used to compute the next initial
voltage value $V_{int}$. 

A subtle issue is created by the additional pure delay $\dmin>0$ in our algorithm. Since
the algorithm runs over all input transitions, the time instants $t^{current}$ and $t^{next}$ refer to input
state transitions. These are also the times where the analog output trajectory
is switched, according to the respective case. However, the time when the
transition of the digital output signal, which is caused by the trajectory
hitting $\vdd/2$, is defered by $\dmin$.
Consequently, the delay functions given in \cref{Theorem:Del_Traj_NOR}, which are used by 
\cref{Alg:NOR} to compute the next output transition time $t_o$ (see e.g. \cref{line:computeto}), 
also incorporate $\dmin$. In addition, the input-to-previous output delay
parameter $T$ refers to the transition time of the previous transition
of the digital output, and \emph{not} to the time where the analog output
voltage crosses $\vdd/2$.
In order to also make sure that $V_{int}^{next}$ for the next input transition
is computed correctly, one needs to add $\dmin$ also to the parameter
$T$ when calling the trajectory formulas provided by \cref{Theorem:Del_Traj_NOR} in \cref{line:computeVinta}, \cref{line:computeVinti}, etc.

Another subtle issue justifies our simple cancellation rules (lines~\cref{line:cancel1}--\cref{line:cancel4}):
The first condition $t^{o} -\dmin < t^{\text{current}}$ in the \textbf{if} statement is true only if the
delay of the input transition at $t^{\text{current}}$ is negative, which characterizes a virtual output
transition that obviously needs to be canceled (but of course remembered). The second condition 
$t^{o} -\dmin> t^{\text{next}}$ is true if the next input transition happens strictly before the 
scheduled output transition. If the next input transition causes the output voltage to change its
direction (rising $\to$ falling or else falling $\to$ rising), the cancellation is mandatory as the
output voltage cannot reach the threshold voltage $\vdd/2$ any more. If the next input transition
does not cause the output voltage to change its direction, it can only cause a next (falling) virtual
transition that happens either before the already scheduled (falling) output transition or else after it.
In either case, it is safe to remove the already scheduled output transition, as the one to be
remembered is the one cause by the input transition at $t^{\text{next}}$.

\medskip

\cref{Theorem:Del_Traj_NOR} below finally provides the delay and trajectory formulas required by \cref{Alg:NOR},
for each of the cases given in \cref{table:case}. The delay formulas $\delta^{\downarrow}_{(a)}(V_{int})$ etc.,
where $\downarrow$ resp.\ $\uparrow$ represent a falling resp.\ rising output transition,
are parametrized by $V_{int}$, which
implicitly depends on the previous-output-to-input delay $T$, and also by $\Delta$ in Case~(g) and Case~(h). 
$V_{int}$ is the output voltage value at the time the current output trajectory (for the given case)
starts, which is given by the voltage value the \emph{previous} output trajectory assumes at the trajectory
switching time (which is also the input switching time). The trajectory formulas $V_{out}^{(a)}(t)$ etc.\ allow \cref{Alg:NOR} to compute the next initial value $V_{int}^{next}=V_{out}(T+\dmin)$, cp.~\cref{line:computeVinta}.

\begin{thm}[Trajectory resp.\ delay formulas for interconnected \NOR\ gates, used in \cref{Alg:NOR}]\label{Theorem:Del_Traj_NOR}
For any $0 \leq |\Delta| \leq \infty$, where $\Delta=t_B-t_A$ is the input separation time of the last
two falling input transitions at or before the input transition that starts the given Case~(a) to (h) 
output trajectory from the initial value $V_{int}$, 
the analytic delay formulas and output trajectories are as follows:
\begin{flushleft}
Case~(a) and Case~(f):
\end{flushleft} 
\begin{align}
\delta^{\downarrow}_{(a/f)}(V_{int})= C_1R_{n_{A}} \log\Bigl(\frac{2V_{int}}{\vdd}\Bigr) + \dmin \label{eq:Gdelay_a}\\
V_{out}^{(a/f)}(t)=\frac{\vdd}{2}e^{\frac{-t}{C_1R_{n_{A}}}}
\end{align}

\begin{flushleft}
Case~(b) and Case~(e):
\end{flushleft} 
\begin{align}
\delta^{\downarrow}_{(b/e)}(V_{int})= C^{'}_{1}R_{n_{B}} \log\Bigl(\frac{2V_{int}}{\vdd}\Bigr) + \dmin \label{eq:Gdelay_b_e} \\
V_{out}^{(b/e)}(t)=\frac{\vdd}{2}e^{\frac{-t}{C^{'}_{1}R_{n_{B}}}} \label{traj_b_e}
\end{align}

\begin{flushleft}
Case~(c) and Case~(d):
\end{flushleft} 
\begin{align}
\delta^{\downarrow}_{(c/d)}(V_{int})= \frac{C_{2}R_{n_{A}}R_{n_{B}} \log\Bigl(\frac{2 V_{int}}{\vdd}\Bigr)}{R_{n_{A}}+ R_{n_{B}}} + \dmin \label{delay_case_c_d}\\
V_{out}^{(c/d)}(t)=\frac{\vdd}{2}e^{\frac{-(R_{n_{A}}+R_{n_{B}})t}{C_{2}R_{n_{A}}R_{n_{B}}}}
\end{align}

\begin{flushleft}
Case~(g):
\end{flushleft}
\begin{align}
&\delta^{\uparrow}_{(g)}(\Delta,V_{int}) \approx 
\begin {cases}
\begin {cases}
\delta_{0}(V_{int}) - \frac{\alpha_1}{\alpha_1+\alpha_2} \Delta + \dmin  &   \ \ 0 \leq \Delta < \frac{(\alpha_1+\alpha_2)(\delta_{0}(V_{int}) - \delta_{\infty}(V_{int}))}{\alpha_1}   \\ 
\delta_{\infty}(V_{int}) + \dmin &   \ \ \Delta \geq \frac{(\alpha_1+\alpha_2)(\delta_{0}(V_{int}) - \delta_{\infty}(V_{int}))}{\alpha_1} 
\end {cases}  &   \ \ V_{int} \leq \frac{\vdd}{2}  \\ 
 - 2RC_3 \log\Bigl(\frac{\vdd}{2( \vdd- V_{int})}\Bigr) + \dmin &   \ \ V_{int} > \frac{\vdd}{2}  \label{eq:Gdelay_g}
\end {cases}
\end{align}
{\scriptsize
\begin{align}
&V_{out}^{(g)}(t,\Delta,V_{int})=\nonumber\\
&\begin{cases}
    \vdd \biggl[ 1-\frac{e^{-t/(2RC_3)}}{2}
        \Bigl(1+\frac{2t}{\,a+\Delta+\sqrt{\chi}+2(\delta^{\uparrow}_{(g)}(\Delta,V_{int})-\dmin)}\Bigr)^{\frac{-A+a}{2RC_3}}
        \Bigl(1+\frac{2t}{\,a+\Delta-\sqrt{\chi}+2(\delta^{\uparrow}_{(g)}(\Delta,V_{int})-\dmin)}\Bigr)^{\frac{A}{2RC_3}}
    \biggr],
    & V_{int}\le\frac{\vdd}{2}
    \\[1.0em]
    \vdd + (V_{int}-\vdd)\,
        e^{-t/(2RC_3)}
        \Bigl(1+\frac{2t}{a+\Delta+\sqrt{\chi}}\Bigr)^{\frac{-A+a}{2RC_3}}
        \Bigl(1+\frac{2t}{a+\Delta-\sqrt{\chi}}\Bigr)^{\frac{A}{2RC_3}},
    & V_{int}>\frac{\vdd}{2}
\end{cases}
\label{traj_g}
\end{align}
}

\begin{flushleft}
Case~(h):
\end{flushleft}

\begin{align}
&\delta^{\uparrow}_{(h)}(\Delta,V_{int}) \approx 
\begin {cases}
\begin {cases}
\delta_{0}(V_{int}) - \frac{\alpha_2}{\alpha_1+\alpha_2} |\Delta| + \dmin  &   \ \ 0 \leq |\Delta| < \frac{(\alpha_1+\alpha_2)(\delta_{0}(V_{int}) - \delta_{-\infty}(V_{int}))}{\alpha_2}   \\ 
\delta_{-\infty}(V_{int}) + \dmin &   \ \ |\Delta| \geq \frac{(\alpha_1+\alpha_2)(\delta_{0}(V_{int}) - \delta_{-\infty}(V_{int}))}{\alpha_2} 
\end {cases}  &   \ \ V_{int} \leq \frac{\vdd}{2}  \\ 
 - 2RC_3 \log\Bigl(\frac{\vdd}{2( \vdd- V_{int})}\Bigr) + \dmin &   \ \ V_{int} > \frac{\vdd}{2}  \label{eq:Gdelay_h}
\end {cases}
\end{align}
{\scriptsize
\begin{align}
&V_{out}^{(h)}(t,\Delta,V_{int})=\nonumber\\
&\begin{cases}
    \vdd \biggl[ 1-\frac{e^{-t/(2RC_3)}}{2}
        \Bigl(1+\frac{2t}{\,a+|\Delta|+\sqrt{\bchi}+2(\delta^{\uparrow}_{(h)}(\Delta,V_{int})-\dmin)}\Bigr)^{\frac{-\bA+a}{2RC_3}}
        \Bigl(1+\frac{2t}{\,a+|\Delta|-\sqrt{\bchi}+2(\delta^{\uparrow}_{(h)}(\Delta,V_{int})-\dmin)}\Bigr)^{\frac{\bA}{2RC_3}}
    \biggr],
    & V_{int}\le\frac{\vdd}{2}
    \\[1.0em]
    \vdd + (V_{int}-\vdd)\,
        e^{-t/(2RC_3)}
        \Bigl(1+\frac{2t}{a+|\Delta|+\sqrt{\bchi}}\Bigr)^{\frac{-\bA+a}{2RC_3}}
        \Bigl(1+\frac{2t}{a+|\Delta|-\sqrt{\bchi}}\Bigr)^{\frac{\bA}{2RC_3}},
    & V_{int}>\frac{\vdd}{2}
\end{cases}
\label{traj_h}
\end{align}

}

Herein,
\begin{align}
\delta_{0}(V_{int}) =- \frac{\alpha_1 + \alpha_2}{2R} \biggl[ 1+ W_{-1}\biggl(\frac{-1}{e \cdot \Bigl(\frac{2(\vdd-V_{int})}{\vdd} \Bigr)^{\frac{4R^2C_3}{\alpha_1+ \alpha_2}}}\biggr) \biggr], \label{exterimal1}\\
\delta_{\infty}(V_{int})= -\frac{\alpha_2}{2R} \biggl[ 1+ W_{-1}\biggl(\frac{-1}{e \cdot \Bigl(\frac{2(\vdd-V_{int})}{\vdd} \Bigr)^{\frac{4R^2C_3}{\alpha_2}}}\biggr) \biggr], \label{exterimal2}\\
\delta_{-\infty}(V_{int})= -\frac{\alpha_1}{2R} \biggl[ 1+ W_{-1}\biggl(\frac{-1}{e \cdot \Bigl(\frac{2(\vdd-V_{int})}{\vdd}\Bigr)^{\frac{4R^2C_3}{\alpha_1}}}\biggr) \biggr], \label{exterimal3}
\end{align}
$C_1$, $C^{'}_{1}$, $C_2$, and $C_3$ are defined by \cref{eq:C_1_int}, \cref{eq:C_1_prime_int}, \cref{eq:C_2_int}, and \cref{eq:C_3_int}, respectively, and $a$, $\chi$ and $A$ as well as $\bchi$ and $\bA$ have been defined in \cref{thm:MISOuttraj}.

\end{thm}

\begin{proof}
According to the truth table of the \NOR\ gate, six out of the total eight transitions (namely, Case~(a)--(f)) in \cref{table:case} cause a falling output, the remaining two (Case~(g) and~(h)) produce a rising output. It suffices to 
consider Case~(a), Case~(c), Case~(e) and Case~(g), which are associated with non-negative values of $\Delta$. 
The remaining Case~(b), Case~(d), Case~(f), and Case~(h) can be determined from Case~(a), Case~(c), Case~(e), and Case~(g) by applying our symmetry rules, namely replacing $R_{n_{A}}, R_{n_{B}}, \alpha_1, \alpha_2$, $A$, $\chi$ and $\Delta$ with $R_{n_{B}}, R_{n_{A}}, \alpha_2, \alpha_1$, $\bA$, $\bchi$ and $|\Delta|$, respectively.

All that needs to be done in Case~(a)--(f) is to replace $V_{out}^{T^{\uparrow_{-}}}(0)$ in \cref{outsig1}, $V_{out}^{T^{\uparrow_{-}}}(\Delta)$ in \cref{outsig2}, $V_{out}^{T^{\downarrow_{-}}}(0)$ in \cref{eq:FirstFalltheorem}, and $V_{out}^{T^{\downarrow_{-}}}(\Delta)$ in \cref{SoughtOutput} by the arbitrary initial voltage $V_{int}$, and then to invert the resulting trajectories to determine the appropriate delay function.
We exemplify this procedure for Case~(a). Our replacement transforms \cref{outsig1} to $V_{out}^{T^{\uparrow}_{-}}(t) =V_{int} e^{\frac{-t}{C_1 R_{n_{A}}}}$, where $V_{int}=V_{out}^{T^{\uparrow}_{-}}(0)$. Inverting this simple exponential function for determining the time it takes for the trajectory to cross the threshold voltage $\vdd/2$ and adding $\dmin$ to cope with the delayed digital output transition leads to \cref{eq:Gdelay_a}. In order to also compute the voltage trajectory value reached in some time interval $t''$ when starting from $\vdd/2$ instead of $V_{int}$, one needs to apply an appropriate time shift. Letting $t'$ denote the time where $V_{out}^{T^{\uparrow}_{-}}(t')=\vdd/2$ when starting from $V_{int}$ at time 0, and $t'' \doteq t-t'$, 
we obtain $V_{int}=\frac{\vdd}{2}e^{\frac{t'}{C_1 R_{n_{A}}}}$ and hence $V_{out}^{T^{\uparrow}_{-}}(t) = V_{out}^{T^{\uparrow}_{-}}(t'+t'') =\frac{\vdd}{2}e^{\frac{- t''}{C_1 R_{n_{A}}}}$. We can hence define the shifted trajectory formula associated with Case~(a) as $V_{out}^{(a)}(t'') \doteq \frac{\vdd}{2}e^{\frac{- t''}{C_1 R_{n_{A}}}}$. Note that $t''\geq 0$ (resp.\ $t'' < 0$) if we want to compute the trajectory voltage value at a time 
$t$ after (resp.\ before) it crossed $\vdd/2$ at time $t'$.

A similar argument applies for Case~(c) and Case~(e), which also involve a simple exponential function only. 

To obtain the delay formula for the more complex Case~(g), we first replace 
$V_{out}^{T^{\downarrow}_{-}}(\Delta)$ by $V_{int}$ in \cref{SoughtOutput}.
Unfortunately, we need to consider two different cases regarding $V_{int}$ here, 
namely, (i) $V_{int} \leq \vdd/2$ and (ii) $V_{int} > \vdd/2$. 

In the case of (i), the trajectory will reach $\vdd/2$ at or after the time when the trajectory
caused by the input transition $(0,1) \rightarrow (0,0)$ has started (from $V_{int}$), 
leading to a (real or virtual) rising output transition.
We hence just need to solve $V_{out}^{T^{\downarrow \downarrow}_{+}}(t)=\vdd/2$, that is,
\begin{align}
 e^{\frac{-t}{2RC_3}} \Bigl(1+\frac{2t}{d+\sqrt{\chi}}\Bigr)^{\frac{-A+a}{2RC_3}} \Bigl(1+\frac{2t}{d-\sqrt{\chi}}\Bigr)^{\frac{A}{2RC_3}} = \frac{\vdd}{2(\vdd- V_{int})}\label{eq:hitting},
\end{align}
for non-negative $t$. Fortunately, this is just \cite[Eq.~(37)]{ferdowsi2025faithful}, with the constant $1/2$ replaced by $\frac{\vdd}{2(\vdd- V_{int})}$.
Literally following the proof of \cite[Lemma~16 and Theorem~6.5]{ferdowsi2025faithful} effortlessly leads to the generalized approximate delay formula given in \cref{eq:Gdelay_g}. More specifically, \cref{eq:Gdelay_g} turns out to be exactly the same as \cite[Eq.~(52)]{ferdowsi2025faithful}, with modified extremal values $\delta_0$, $\delta_\infty$, and $\delta_{-\infty}$, as defined in \cref{exterimal1}, \cref{exterimal2}, and \cref{exterimal3}.
Note that $V_{int}$ must indeed be less than or equal to $\vdd/2$ here, to ensure that the argument of the Lambert $W_{-1}$ function in \cref{exterimal1}, \cref{exterimal2}, and \cref{exterimal3} is larger than or equal to $-1/e$. 

Applying the time shift to the corresponding trajectory, i.e., to \cref{SoughtOutput}, is more tedious. For computing the voltage trajectory value $V_{out}^{T^{\downarrow \downarrow}_{+}}(t)$ reached at some time $t''$ after 
it crossed $\vdd/2$, we again
consider the time shift $t''=t-t'$, where $t'$ is the time where $V_{out}^{T^{\downarrow \downarrow}_{+}}(t')=\vdd/2$, i.e., the solution of \cref{eq:hitting}. Rearranging \cref{eq:hitting} easily allows to compute the value of $V_{int}$ that leads to the solution $t'$:
\begin{align}
V_{int}= \vdd \Bigl( 1- \frac{1}{2e^{\frac{-t'}{2RC_3}} \bigl(1+\frac{2t'}{d+\sqrt{\chi}}\bigr)^{\frac{-A+a}{2RC_3}} \bigl(1+\frac{2t'}{d-\sqrt{\chi}}\bigr)^{\frac{A}{2RC_3}}} \Bigr) \label{Eq:Vint}.
\end{align}
On the other hand, incorporating our time shift in \cref{SoughtOutput} yields
{\footnotesize
\begin{align}
&V_{out}^{T^{\downarrow \downarrow}_{+}}(t)= V_{out}^{T^{\downarrow \downarrow}_{+}}(t''+t')= \vdd  + \bigl(V_{int} -\vdd   \bigr)  \cdot 
  \left[ e^{\frac{-t'}{2RC_3}} \cdot e^{\frac{-t''}{2RC_3}} \Bigl(1+\frac{2(t''+t')}{d+\sqrt{\chi}}\Bigr)^{\frac{-A+a}{2RC_3}} \Bigl(1+\frac{2(t''+t')}{d-\sqrt{\chi}}\Bigr)^{\frac{A}{2RC_3}} \right], \label{ShiftTraj1}
\end{align}
}
which, by plugging in \cref{Eq:Vint}, leads to
\begin{align}
& V_{out}^{(g)}(t'',\Delta,T) = V_{out}^{T^{\downarrow \downarrow}_{+}}(t''+t') = \vdd + \nonumber \\
& \qquad\qquad\biggl( \vdd \Bigl( 1- \frac{1}{2e^{\frac{-t'}{2RC_3}} \bigl(1+\frac{2t'}{d+\sqrt{\chi}}\bigr)^{\frac{-A+a}{2RC_3}} \bigl(1+\frac{2t'}{d-\sqrt{\chi}}\bigr)^{\frac{A}{2RC_3}}}\Bigr) -\vdd \biggr) \cdot \nonumber \\
& \qquad\qquad\qquad\biggl[ e^{\frac{-t'}{2RC_3}} \cdot e^{\frac{-t''}{2RC_3}} \Bigl(1+\frac{2(t''+t')}{d+\sqrt{\chi}}\Bigr)^{\frac{-A+a}{2RC_3}} \Bigl(1+\frac{2(t''+t')}{d-\sqrt{\chi}}\Bigr)^{\frac{A}{2RC_3}}  \bigr) \biggr] =\nonumber \\
&\vdd \biggl[ 1- \frac{e^{\frac{-t'}{2RC_3}} e^{\frac{-t''}{2RC_3}} \Bigl(1+\frac{2(t''+t')}{d+\sqrt{\chi}}\Bigr)^{\frac{-A+a}{2RC_3}} \Bigl(1+\frac{2(t''+t')}{d-\sqrt{\chi}}\Bigr)^{\frac{A}{2RC_3}} \Bigr)}{2e^{\frac{-t'}{2RC_3}} \Bigl(1+\frac{2t'}{d+\sqrt{\chi}}\Bigr)^{\frac{-A+a}{2RC_3}} \Bigl(1+\frac{2t'}{d-\sqrt{\chi}}\Bigr)^{\frac{A}{2RC_3}}}  \biggr]= \nonumber \\
&\vdd \biggl[ 1- \frac{1}{2} e^{\frac{-t''}{2RC_3}} \Bigl(1+\frac{2t''}{d+\sqrt{\chi}+2t'}\Bigr)^{\frac{-A+a}{2RC_3}} \Bigl(1+\frac{2t''}{d-\sqrt{\chi}+2t'}\Bigr)^{\frac{A}{2RC_3}} \biggr].\label{eq:shiftedtraj}
\end{align}

Whereas the employed time shift has not caused $t'$ to cancel out completely in \cref{eq:shiftedtraj}, recalling that $t'\geq 0$ is the solution of \cref{eq:hitting}, which is actually $\delta^{\uparrow}_{(g)}(\Delta,V_{int})-\dmin$ as computed in  \cref{eq:Gdelay_g}, and that $d=a+ \Delta$ according to \cref{thm:MISOuttraj} finally gives

{\footnotesize
\begin{flalign}
&V_{out}^{(g)}(t'',\Delta,V_{int}) =\nonumber\\ &
\vdd \biggl [ 1-\frac{e^{\frac{-t''}{2RC_3}}}{2} \Bigl(1+ \frac{2t''}{a+ \Delta+ 
\sqrt{\chi}+2 (\delta^{\uparrow}_{(g)}(\Delta,V_{int})-\dmin)}\Bigr)^{\frac{-A+a}{2RC_3}} \Bigl(1+ \frac{2t''}{a+ \Delta- \sqrt{\chi}+ 2 (\delta^{\uparrow}_{(g)}(\Delta,V_{int})-\dmin)}
\Bigr)^{{\frac{A}{2RC_3}}} \biggr] \nonumber &
\end{flalign}}
and hence \cref{traj_g}.

\medskip

Unfortunately, the above reasoning does not apply for the other case (ii), where $V_{int}>\vdd/2$.
More specifically, the trajectory must have hit $\vdd/2$ already before the time where it started 
off from $V_{int}$ here. Still, \cref{eq:hitting} does not have a solution for negative values of $t$, i.e., our 
complex trajectory formula governing Case~(g) is not sufficiently well-behaved when going back in 
time \emph{before} the initial state. Fortunately, however, nothing prevents us from re-defining 
our trajectory formula for negative times in a convenient way: After all, it is only needed by \cref{Alg:NOR} to compute $V_{int}^{next}$ at the time of the next input transition $t^{next}\geq t$, given trajectory~(g)'s 
current $V_{int}$ (assumed at some time $t>t'$) and the time difference $T+\dmin=t^{next}-t'>0$ since the time $t'$ when the output trajectory has has
hit $\vdd/2$. 

We will define our artificial negative trajectory by dropping the two complex factors within the brackets
of \cref{SoughtOutput}, which results in the simple exponential function
\[
\hat{V}_{out}^{T^{\downarrow \downarrow}_{+}}(t)= \vdd  + \bigl(V_{out}^{T^{\downarrow}_{-}}(\Delta) -\vdd   \bigr) \cdot e^{\frac{-t}{2RC_3}}.
\]
Note carefully that it fullfills the required continuity at $t=0$ since $\hat{V}_{out}^{T^{\downarrow \downarrow}_{+}}(0) = V_{out}^{T^{\downarrow \downarrow}_{+}}(0)$. Replacing $V_{out}^{T^{\downarrow}_{-}}(\Delta)$ by $V_{int}$
leads to
\begin{align}
\hat{V}_{out}(t)=\vdd + (V_{int}- \vdd) e^{-t/(2RC_3)}
\label{auxiliaryfunction}
\end{align}
and thus to the analog of \cref{eq:hitting} given by
\begin{align}
 e^{\frac{-t}{2RC_3}} = \frac{\vdd}{2(\vdd- V_{int})}\label{eq:hittinghat},
\end{align}
which does have a (negative) solution $t< 0$ also when $\vdd > V_{int} > \vdd/2$. The $V_{int}$ that causes
the trajectory to hit $\vdd/2$ at time $t'$ is hence
\[
V_{int}= \vdd \Bigl( 1- \frac{1}{2}e^{\frac{t'}{2RC_3}}\Bigr).
\]
Plugging this into \cref{auxiliaryfunction} gives the shifted trajectory $\hat{V}_{out}^{(g)}(t'') \doteq 
\hat{V}_{out}(t'+t'') = \vdd (1- 0.5 e^{-t''/(2RC_3)})$. The corresponding delay, defined as the time at which \cref{auxiliaryfunction} intersected $\vdd/2$ when starting from $V_{it}$ at time 0, is just the solution $\hat{\delta}_{(g)}^{\uparrow}(V_{int}) = -2RC_3 \log \bigl(\vdd /(2(\vdd-V_{int})) \bigr) < 0$ of \cref{eq:hittinghat}. Given $t'$ where the trajectory has hit $\vdd/2$, $V_{int}$, and $\hat{\delta}_{(g)}^{\uparrow}(V_{int})$, one can hence determine the time $t' -\hat{\delta}_{(g)}^{\uparrow}(V_{int})$ where trajectory~(g) has started from $V_{int}$. Consequently, the sought 
$V_{out}^{(g)}(t,\Delta,V_{int})$  for case (ii) is given by \cref{traj_g}, which is just \cref{SoughtOutput} with 
$V_{out}^{T^{\downarrow}_{-}}(\Delta)$ replaced by $V_{int} > \vdd/2$ evaluated at $t=T+\dmin-2RC_3 \log \bigl(\vdd /(2(\vdd-V_{int})) \bigr)$ (cp.~\cref{line:caseg} in \cref{Alg:NOR}). The above construction is illustrated in \cref{fig:Traverse_traj}.

\begin{figure}[t]
  \centerline{
    \includegraphics[width=0.48\linewidth]{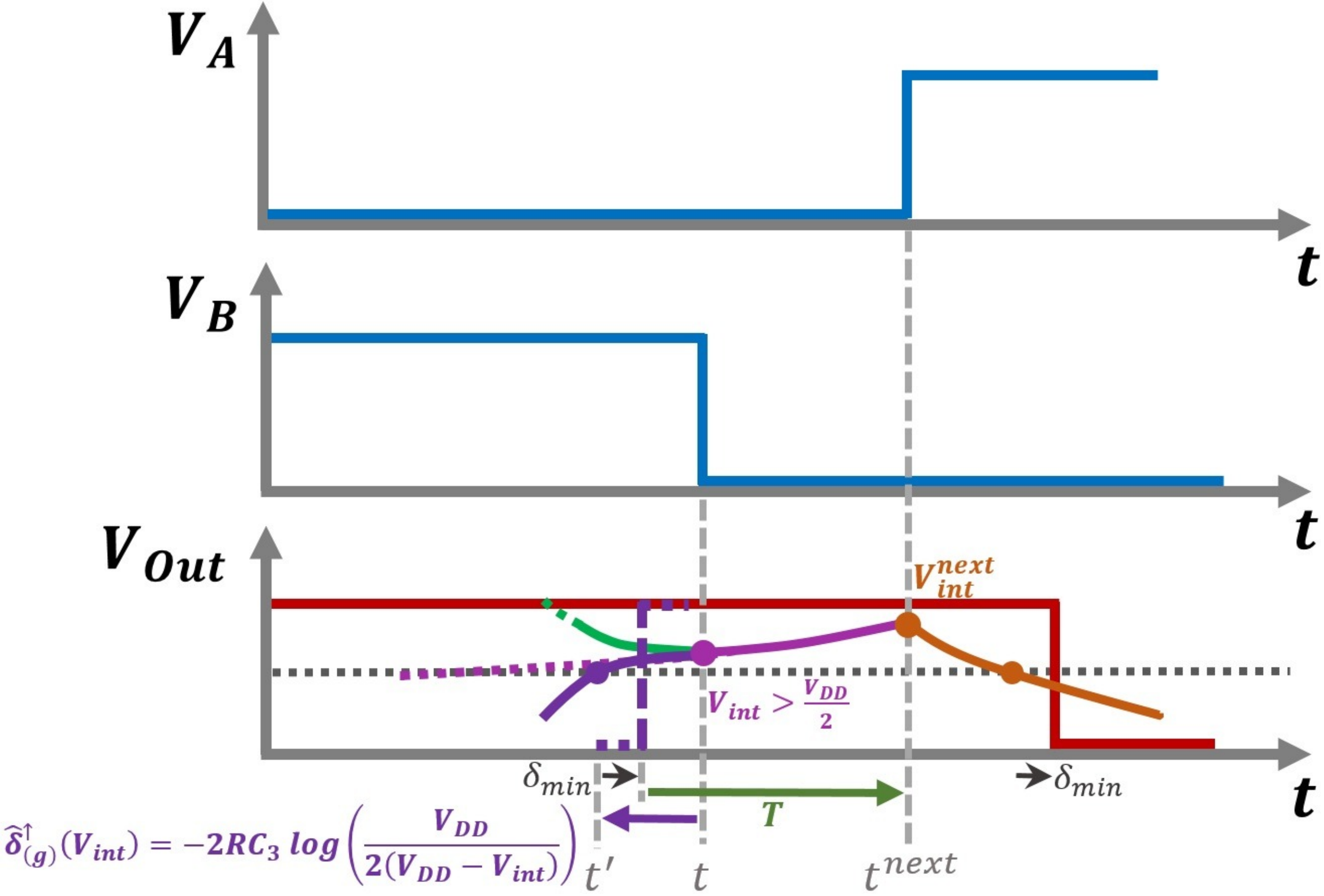}
  }
  \caption{Timing diagram of the \NOR\ gate associated with Case~(g) for $V_{int}>\vdd/2$. Herein, $T-2RC \log \bigl(\vdd /(2(\vdd-V_{int})) \bigr) + \dmin$ specifies the time span its trajectory (pink one) needs to be traversed to compute $V^{next}_{int}$, the desired initial voltage value at the time the next transition happens. The purple trajectory represents the virtual negative exponential trajectory.}
  \label{fig:Traverse_traj}
\end{figure}

We conclude our proof by noting that such artificial negative trajectory functions are only needed for Cases~(g) and~(h). In all other cases, the original trajectories can be arbitrarily shifted, such that the initial value $V_{int}$ can be determined directly by evaluating the appropriately shifted output trajectory at $T+\dmin$.
\end{proof}

\subsection{A full delay model for the 2-input \NAND\ gate}

\begin{figure}[h]
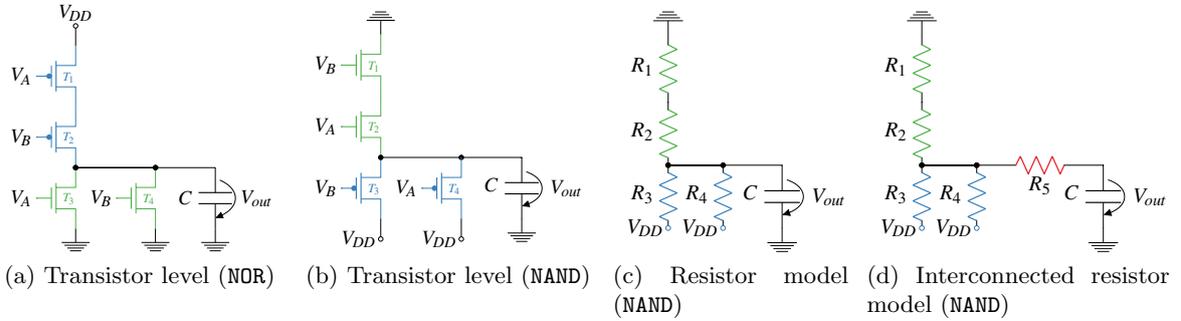

  \centering
  \subfloat[Transistor level (\NOR)]{
\includegraphics[height=0.21\linewidth]{\figPath{nor_RC_colored.pdf}}  
    \label{fig:nor_CMOS2}}
  \hfil
  \subfloat[Transistor level (\NAND)]{
\includegraphics[height=0.2\linewidth]{\figPath{NAND_RC_colored.pdf}}  
    \label{fig:nand_CMOS}}
  \hfil
  \subfloat[Resistor model (\NAND)]{
 \includegraphics[height=0.2\linewidth]{\figPath{nand_R_colored.pdf}}%
    \label{Figurenand-GATE}}
  \hfil
  \subfloat[Interconnected resistor model (\NAND)]{
 \includegraphics[height=0.2\linewidth]{\figPath{nand_R_interconnect2.pdf}}%
    \label{Figurenand-GATE_Int}}    
  \caption{Transistor schematic and the resistor model of a CMOS \NAND\ gate along with its augmented RC interconnect component. In contrast to the \NOR\ gate, the \NAND\ gate uses two series-connected nMOS transistors (partly modeled as time-varying resistors) and two parallel-connected pMOS transistors (modeled as ideal switches).} \label{fig:NANDNORCom}
\end{figure}

In this section, we will provide a full delay model for interconnected 2-input
\NAND\ gates. Thanks to the structural similarity of the CMOS implementation
of a \NAND\ gate and a \NOR\ gate, just compare \cref{fig:nor_CMOS2} and
\cref{fig:nand_CMOS}, this can be accomplished by a simple reduction to the
full delay model of the \NOR\ gate. More specifically, there is a one-to-one
mapping of the possible 2-input-transition cases for the \NAND\ to the
\NOR, as given in \cref{table:case_NAND}.

To justify why such a mapping leads to a full delay model for the \NAND\
gate with two inputs $A$ and $B$, we use De Morgan's law
to get the logical equivalence  $\neg (A \land B) =
\neg \bigl( \neg ( \neg A \lor \neg B) \bigr)$. A
\NAND\ gate can hence be implemented by first (i) feeding negated
inputs into a \NOR\ gate, followed by (ii) negating the output.
Viewed from the perspective of a CMOS implementation, (i)
can be accomplished by replacing pMOS transistors with nMOS
transistors and vice versa in the \NOR\ implementation
\cref{fig:nor_CMOS2}, and (ii)
by swapping $\vdd$ with ground, which indeed leads to the \NAND\
implementation \cref{fig:nand_CMOS}.

To make the variable resistor
model introducted in \cref{sec:Prel} applicable, we just assume
that the serial nMOS transistors exhibit a time-varying resistance
when switching-on, whereas their switching-off, as well as any
switching of the parallel pMOS transistors happens in zero
time. \cref{Figurenand-GATE} resp.\ \cref{Figurenand-GATE_Int} shows
the resulting resistor models without resp.\ with the additional
resistor $R_5$ for the interconnect.

The above discussion justifies the 1:1 correspondence between the different
2-input scenarios for the \NAND\ gate and those for the \NOR\ gate given
in \cref{table:case_NAND}. For every possible input transition of a
\NAND\ gate, it specifies those transition of a \NOR\ gate where our
variable resistor model behaves identically. The underlying mapping is
achieved by negating the inputs and identifying the corresponding case in \cref{table:case}. For instance, to determine the shifted trajectory and delay formula for the transition $(0,0) \rightarrow (1,0)$ in a \NAND\ gate (i.e., Case (a)), we first consider its negated equivalent in the \NOR\ gate, which is $(1,1) \rightarrow (0,1)$. According to \cref{table:case}, this transition corresponds to Case (e). Therefore, the appropriate delay formula and the shifted voltage trajectory for Case (a) in the \NAND\ gate can be derived from \cref{eq:Gdelay_b_e} and \cref{traj_b_e}.

\begin{table}[h]
\centering
\caption{\small\em Different cases associated with varying transitions of inputs of a \NAND\ gate and their corresponding \NOR\ cases.}
\label{table:case_NAND}
\scalebox{0.9}{
\begin{tabular}{lcccccccc}
\hline
\multicolumn{4}{c}{\NAND\ gate}                                                 & \multicolumn{1}{l}{} & \multicolumn{4}{c}{Corresponding \NOR\ gate}                            \\ \cline{1-4} \cline{6-9} 
\multicolumn{1}{c}{Case} & Notation            & State transition     & Output       &                      & Case & Notation            & State transition     & Output       \\ \cline{1-4} \cline{6-9} 
(a)                      & $(\uparrow,0-,t)$   & $(0,0) \rightarrow (1,0)$   & 1  &                    & (e)  & $(\downarrow,1-,t)$ & $ (1,1) \rightarrow (0,1)$ & 0  \\
(b)                      & $(0-,\uparrow,t)$   & $(0,0) \rightarrow (0,1)$  & 1  &                      & (f)  & $(1-,\downarrow,t)$ & $ (1,1) \rightarrow (1,0)$ & 0 \\
(c)                      & $(1-,\uparrow,t)$   & $(1,0) \rightarrow (1,1)$   & 0  &                     & (g)  & $(0-,\downarrow,t)$ & $ (0,1) \rightarrow (0,0)$ & 1 \\
(d)                      & $(\uparrow,1-,t)$   & $ (0,1) \rightarrow (1,1)$  & 0 &                      & (h)  & $(\downarrow,0-,t)$ & $ (1,0) \rightarrow (0,0)$ & 1 \\
(e)                      & $(\downarrow,1-,t)$ & $ (1,1) \rightarrow (0,1) $ & 1 &                     & (a)  & $(\uparrow,0-,t)$   & $(0,0) \rightarrow (1,0)$   & 0\\
(f)                      & $(1-,\downarrow,t)$ & $ (1,1) \rightarrow (1,0)$  & 1 &                     & (b)  & $(0-,\uparrow,t)$   & $(0,0) \rightarrow (0,1)$  & 0 \\
(g)                      & $(0-,\downarrow,t)$ & $ (0,1) \rightarrow (0,0)$  & 1 &                     & (c)  & $(1-,\uparrow,t)$   & $(1,0) \rightarrow (1,1)$   & 0 \\
(h)                      & $(\downarrow,0-,t)$ & $ (1,0) \rightarrow (0,0)$  & 1 &                     & (d)  & $(\uparrow,1-,t)$   & $ (0,1) \rightarrow (1,1)$ & 0 \\ \hline
\end{tabular}}
\end{table}

However, an additional adjustment is required to obtain the correct trajectory formulas for the \NAND\ gate from the trajectory formulas for the \NOR\ gate: As previously mentioned, in addition to negating the inputs, the output of the \NOR\ gate must also be negated. At the level of voltage trajectories, this is accomplished by just subtracting the \NOR\ gate trajectories from $\vdd$. Note carefully that this also
requires to instantiate the corresponding \NOR\ gate trajectory with $\vdd-\vint$. 
As a concrete example, consider Case~(a) and Case~(c) for the \NAND\ gate. To determine the voltage trajectories for these two cases, we need to use Case~(e) and Case~(g) of the \NOR\ gate (i.e., \cref{traj_b_e} and \cref{traj_g}, respectively). The sought \NAND\ gate trajectories are obtained by subtracting these expressions (after replacing $\vdd$ by $\vdd-\vint$) from $\vdd$, yielding
\begin{align}
V_{out\_ NAND}^{(a)}(t)= \vdd \bigl( 1-0.5 e^{\frac{-t}{C^{'}_{1}R_{n_{B}}}}\bigr)  \label{traj_e_NAND}
\end{align}
and
{ \scriptsize
\begin{align}
&V_{out\_NAND}^{(c)}(t,\Delta,V_{int})=\nonumber\\ 
&\begin{cases}
    \vdd \biggl[ \frac{e^{-t/(2RC_3)}}{2}
        \Bigl(1+\frac{2t}{\,a+\Delta+\sqrt{\chi}+2(\delta^{\uparrow}_{(g)}(\Delta,\vdd-V_{int})-\dmin)}\Bigr)^{\frac{-A+a}{2RC_3}}
        \Bigl(1+\frac{2t}{\,a+\Delta-\sqrt{\chi}+2(\delta^{\uparrow}_{(g)}(\Delta,\vdd-V_{int})-\dmin)}\Bigr)^{\frac{A}{2RC_3}}
    \biggr],
    & V_{int}>\frac{\vdd}{2}
    \\[1.0em]
    \bigl(\vdd - V_{int}\bigr)\,
        e^{-t/(2RC_3)}
        \Bigl(1+\frac{2t}{a+\Delta+\sqrt{\chi}}\Bigr)^{\frac{-A+a}{2RC_3}}
        \Bigl(1+\frac{2t}{a+\Delta-\sqrt{\chi}}\Bigr)^{\frac{A}{2RC_3}},
    & V_{int}\le\frac{\vdd}{2}
\end{cases}
\label{traj_g_NAND}
\end{align}

}

Following an analogous argument allows to determine the delay formulas for the \NAND\ gate. Note that
it is instrumental here that our threshold crossing voltage is $\vth=\vdd/2$. For Case~(a) and Case~(c),
for example, \cref{eq:Gdelay_b_e} and \cref{eq:Gdelay_g} effortlessly lead to
\begin{align}
\delta^{\uparrow}_{(a\_NAND)}(V_{int})= C^{'}_{1}R_{n_{B}} \log\Bigl(\frac{2 (\vdd- V_{int})}{\vdd}\Bigr) + \dmin \label{eq:GdelayNAND_a}
\end{align}

{\scriptsize
\begin{align}
&\delta^{\downarrow}_{(c\_NAND)}(\Delta,V_{int}) \approx 
\begin {cases}
\begin {cases}
\delta_{0}(\vdd- V_{int}) - \frac{\alpha_1}{\alpha_1+\alpha_2} \Delta + \dmin  &    0 \leq \Delta < \frac{(\alpha_1+\alpha_2)(\delta_{0}(\vdd-V_{int}) - \delta_{\infty}(\vdd-V_{int}))}{\alpha_1}   \\ 
\delta_{\infty}(\vdd-V_{int}) + \dmin &   \Delta \geq \frac{(\alpha_1+\alpha_2)(\delta_{0}(\vdd-V_{int}) - \delta_{\infty}(\vdd-V_{int}))}{\alpha_1} 
\end {cases}  &    V_{int} > \frac{\vdd}{2}  \\ 
 - 2RC_3 \log\Bigl(\frac{\vdd}{2V_{int}}\Bigr) + \dmin &   V_{int} \leq \frac{\vdd}{2}  \label{eq:GdelayNAND_c}
\end {cases}
\end{align}
}
with, $\delta_{0}$, $\delta_{\infty}$, and $\delta_{-\infty}$ defined in \cref{exterimal1}, \cref{exterimal2}, and \cref{exterimal3}.

\subsection{Modeling other gates}

A crucial feature of a delay modeling approach like ours is applicability to different types of
gates. Besides the \NAND\ gate, our approach can
be easily applied to every gate that involves serial or parallel transistors, like Muller \cg\ and 
\AOIgate\ (and-or-inverter). Note that the main effort here is the development of the analytic 
delay formulas, which needs to be done only once, and not the computational costs when using these 
formulas in dynamic timing analysis (which are essentially the same for any gate). For the 2-input Muller \cg\ gate, this has been done in \cite{ferdowsi2024hybrid} already.

On the other hand, increasing the fan-in inevitably requires a substantial effort.
First and foremost, this causes the number of different MIS cases (and
hence of $\Delta$-values) to increase exponentially, which is
inevitable for any MIS-aware delay model, cp.~e.g.\ \cite{CS96:DAC,CGB01:DAC,SRC15:TODAES,AKMK06:DAC}. 
For example, for
a 3-input \NOR\ gate starting from input $A=0$, $B=0$, $C=0$, the
time for the output to go from 1 to 0 depends on both the time 
differences $\Delta_{AB}=t_B-t_A$ and $\Delta_{BC}=t_C-t_B$., where
$t_A$, $T_B$, $t_C$ denote the falling transition times of inputs
$A$, $B$, $C$. This not only substantially increases the number of 
different delay formulas needed for describing a single gate,
but also requires a refined trace history analysis for selecting
the delay formula to be used for some specific output transition
in dynamic timing analysis. 

More importantly, however, increasing the fan-in also substantially
complicates the development of the analytic delay formulas: Whereas
this does not increase the order of the ODEs in the underlying 
hybrid model, it does increase the order of the rational functions arising
in the integrals shown in \cref{T:InerInt}. This does not prohibit 
a closed-form solution of the resulting integrals, the additional zeros 
in the denomiators of the solutions will certainly complicate the development 
of the explicit trajectory formulas and, in particular, their inverses, however.

\section{Implementation in the Involution Tool}
\label{sec:idm_sim}

The Involution Tool (\invt), described in detail in \cite{OMFS20:INTEGRATION,ohlinger2018involution}, is an open-source tool for digital dynamic timing analysis based on the state-of-the-art simulation tool Questa. Its unique asset is a full implementation of the IDM \cite{FNNS19:TCAD} for single-input single-output delay channels, which are used to
add delays to zero-time boolean logic gates. In a simulation run,
for every input transition occurring at such a channel, the \invt\ calculates the next transition time $T$, determines $\delta(T)$, and schedules the next channel output transition accordingly.

Overall, the Involution Tool mirrors VHDL Vital, in the sense that it has
identical structure and variables, is also written in VHDL, and is completely controlled by Questa. However, it utilizes the IDM in its simulation runs, whereas VHDL Vital uses static 
delay values only.
Since switching between VHDL Vital and the Involution Tool just amounts to
switching libraries, the latter inherently facilitates full code 
and test-setup re-use and provides all Questa simulation features as well:
For \invt\ simulations, the same input files are necessary as with any standard post-layout simulation, including the circuit and a testbench. Circuit specifications consist of 
a \emph{Spice}/\emph{Verilog} description and a timing file, which can be created with \emph{Cadence Encounter} \cite{Encounter_tutorial2016}, for example. For simulations under the
IDM, they must be complemented by the IDM model parameters in \emph{.sdf} files for every gate of the circuit.

\subsection{Main features of the Involution Tool}
\label{sec:IT}

\begin{figure}[t!]
  \centering
  \includegraphics[width=0.37\columnwidth]{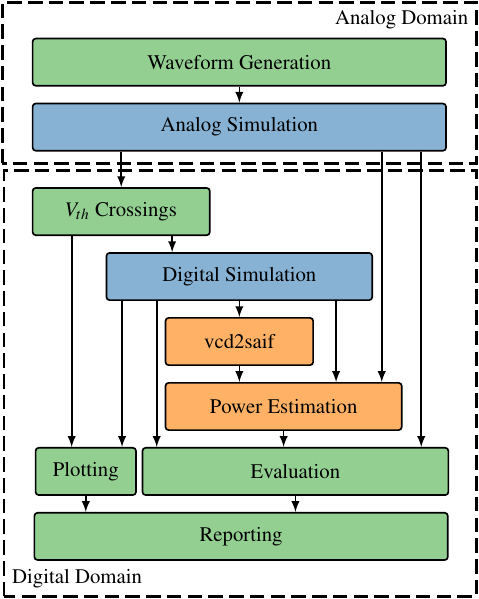}
  \caption{Workflow in the \invt . The green parts had to be implemented from
    scratch. The blue parts have been available, albeit the available resources
    had to be extended significantly. Orange parts could be used almost out of
    the box. 
}
  \label{fig:workflow}
\end{figure}

Going way beyond DDTA, the Involution Tool actually provides a whole framework for evaluating different delay prediction methods across various power and timing metrics. 
The overall architecture, illustrated in \cref{fig:workflow}, centers around the state-of-the-art discrete-event digital simulation tool Questa.
However, the \invt\ also supports the generation of user-defined random input vectors, the execution of analog/digital simulations, parameter sweeps, and detailed reporting on the outcomes. The \invt's modular design allows easy substitution of its components, a crucial feature when encountering numerical issues with \hspice, leading to a switch to \emph{Spectre} for certain scenarios.

\subsubsection*{Feature overview}
\paragraph*{Digital stimuli generation:}
The \invt\ allows to generate test stimuli, typically through a user-configured Gaussian distribution determining the transition times at each input. Both synchronized and unsynchronized (local) transition generation is
supported. This process, customizable via a Python script, aims at emulating real signal behaviors, especially for multi-input gates by manipulating input grouping and transition probabilities.

\paragraph*{Analog simulation:}
Analog simulations provide a reference for power consumption and actual switching behaviors at circuit nodes. These simulations use input signals generated from the digital test stimuli by steep analog input transitions shaped via inverter chains to suit specific technological needs, and also facilitates accurate power measurement of the main circuit only.

\paragraph*{Digital simulation:}
All digital simulations leverage Questa, which processes the digital input signals generated from the analog ones. The user can choose between standard VHDL Vital libraries and 
IDM libraries for simulation, provided every gate in the test circuit comes with its corresponding model parameter \emph{.sdf} file.

\paragraph*{Power estimation:}
Besides analog-based power estimation, digital tools like \emph{Design Compiler} and \emph{PrimeTime} are employed, offering different estimation modes. These simulations use switching activity files (.saif) and value change dump files (.vcd) to compare with analog power estimations, providing insights into potential deviations and biases introduced by digital estimation methods.

\paragraph*{Evaluation:}
This feature consolidates simulation results into a unified format, analyzing power deviations, transition counts, and the timing of transitions using normalized area metrics under various conditions, aiding in identifying potential biases or inaccuracies in delay predictions.

\paragraph*{Reporting:}
Finally, the \invt\ compiles a comprehensive \LaTeX\ report detailing the simulation settings, results, and comparisons, with extensive customization options available through a templating system.




\subsubsection*{Evaluation data output}
For evaluating the accuracy of a model, the \invt\ provides
the deviation between a digital simulation trace generated by
some model and the switching trace generated from an analog
simulation using the same inputs. More specifically, it computes
the \emph{area under the deviation trace}, with and without induced resp.\
suppressed glitches: It accumulates signed areas, where the sign of a 
contributing area depends on whether the
  transition of the reference signal comes first (negative, representing a
  \emph{trailing} transition) or not (positive, representing a \emph{leading}
  transition). The accumulated area is then normalized to
  $V_{DD}=1$ and to the number of transitions, i.e., divided by the number of relevant
transitions. It therefore effectively represents the
  (average) time a transition happens before or after the reference signal.

During the comparison of a \spice-generated switching trace and the corresponding
digital simulation trace, the \invt\ also
  checks whether a pulse (= two subsequent transitions, starting from and returning to 
the current level of the other trace) happens in one trace without any transition
  in the other one:  If such a pulse occurs in the \spice\ switching trace, it is called
an  \emph{original suppressed glitch}, otherwise an \emph{original induced glitch}. 
The \invt\ also determines whether a pulse in one trace properly contains a pulse in the other trace; 
such a pulse is called an
  \emph{inverted suppressed glitch} resp.\ an \emph{inverted induced glitch}.
Besides optionally incorporating those glitches in the deviation area computations, the \invt\ can also output the number of those glitches divided by the total number of 
transitions in the corresponding signal.

\begin{figure}[ht]
	\centering
	\includegraphics[width=0.6\columnwidth]{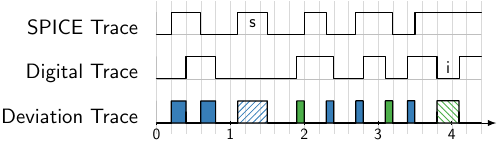}
	\caption{Categorization of the deviation between analog and digital simulation. Shown are pulses (dashed areas) caused by original induced (i) or original suppressed 
	glitches (s) and the leading (blue) and trailing metric (green). [Taken from \cite{OMFS20:INTEGRATION}]}
	\label{fig:metric_explanation}
\end{figure}

  \cref{fig:metric_explanation} shows an example trace with leading (blue) and 
  trailing (green) transitions, which also contains a
  suppressed glitch (s) on the \spice\ trace and an induced glitch (i) on the digital trace. 

Note that some caveat is in order when comparing numerical values of the leading
(or trailing) metric per transition for different models: 
Even if the underlying
\spice\ trace is the same, the \emph{number}
of relevant transitions for different
models is likely to be different, due to suppressed and/or induced glitches. 
Since such glitches barely change the total area under the
deviation trace, the different numbers of transitions may lead to a possibly
significantly different normalized area per transition.

\subsection{The Involution Tool extension \cite{OS23:DDECS}}

An important feature of the Involution Tool is easy extendability.
For example, all that is needed to incorporate
additional gates is to model their Boolean functionality in
VHDL, and linking inputs and outputs through appropriate IDM
channels. Adding additional IDM channels just requires to implement
their delay functions in C and link it to Questa.

Adding support for multi-input gates like 2-input \NOR, \NAND\ and \cg\
gates is more complex, however, since they are not just single-input single-output
channels. Fortunately, the extension of the Involution Tool developed for
the $\eta$-CIDM in \cite{OS23:DDECS} greatly simplifies this problem,
since it not only provided the basic support for multi-input gates but
also an interface that allows to implement the simulation algorithm and
the delay functions in Python.
Implementing \cref{Alg:NOR} and the formulas given in \cref{Theorem:Del_Traj_NOR}
was hence straightforward.

\section{Simulation results}
\label{Sec:experiments}
\begin{figure}[h]
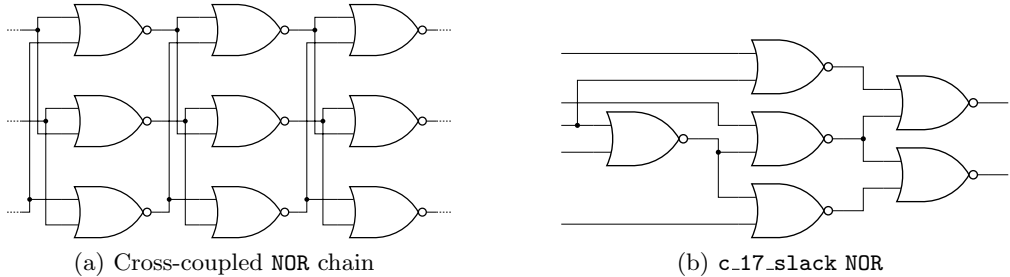

  \centering
  \subfloat[Cross-coupled \NOR\ chain]{
    \includegraphics[width=0.37\linewidth]{\figPath{Involution_tool_benchmarking_circuit.pdf}}%
    \label{fig:InvBench}}
  \hfil
  \subfloat[\texttt{c\_17\_slack} \NOR]{
    \includegraphics[width=0.37\linewidth]{\figPath{c17_slack.pdf}}%
    \label{fig:c17_slack}}
  \caption{ \small $(a)$ Illustration of three stages of our cross-coupled \NOR\ chain. $(b)$ Schematics of the 
\texttt{c\_17\_slack} benchmark circuit, where all \NAND\ gates have been replaced by \NOR\ gates.}\label{fig:Experiment}
\end{figure}

In order to showcase the accuracy improvement of our new model, we use two
different target circuits. The first one consists of three parallel chains 
of 50\footnote{Actually, it also contains 5 additional stages for input pulse 
shaping.} identical cross-coupled \NOR\ gates, as shown in \cref{fig:InvBench}.
We synthesized this circuit
using the Nangate Open Cell Library with FreePDK15$^\text{TM}$
\SI{15}{\nm} FinFET models \cite{Nangate15} ($\vdd=\SI{0.8}{\V}$) and 
measured both the MIS delays and the pure delay $\dmin$ 
of the constituent \NOR\ gates. For determining
$\dmin$, we used the method proposed in Maier et~al.\ in \cite{maier2021composable}
(which abstracts away the delay caused by the slope of the analog waveforms)
after tying together the inputs $A$ and $B$ of our \NOR\ gate. For the extracted
load capacitance $C$, we used the model parametrization given in {\cite[Theorem 3]{ferdowsi2024hybrid}} for computing the model parameters. The resulting values are summarized in 
\cref{table:ParamNORchain}.

\begin{figure}[ht]
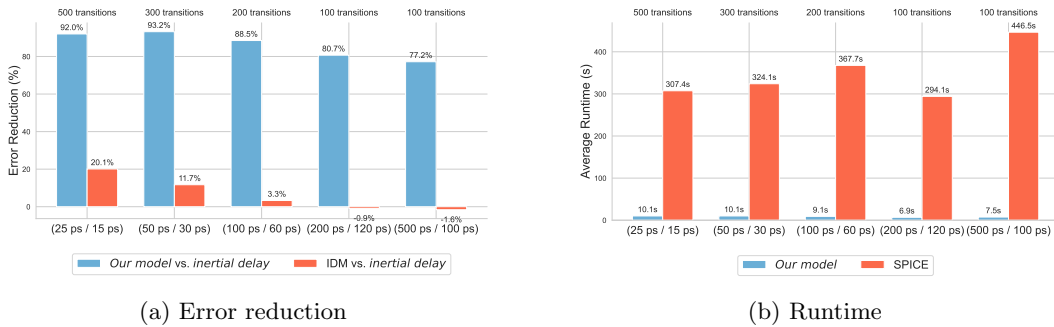

  \centering
  \subfloat[Error reduction]{
    \includegraphics[width=0.41\linewidth]{\figPath{NOR_chain_accuracy}}%
    \label{fig:NORPar_acc}}
  \hfil
  \subfloat[Runtime]{
    \includegraphics[width=0.41\linewidth]{\figPath{NOR_chain_Runtime}}%
    \label{fig:NORPar_TC}}
  \caption{ \small
  $(a)$: Average error reduction of (i) our model vs.\ inertial delay and (ii) the IDM vs.\ inertial delays for cross-coupled \NOR\ chain. $(b)$: Comparison of average runtime of \spice\ and our model for cross-coupled \NOR\ chain.
  }\label{fig:Simresults}
\end{figure}

\begin{table}[ht]
\centering
\caption{\NOR\ model parameter values for the \NOR\ chain circuit.}
\label{table:ParamNORchain}
\scalebox{0.7}{
\begin{tabular}{|clcc|}
\hline
\multicolumn{2}{|c|}{$\dupD_S(-\infty)=3.3798 \ \si{ps}$} & \multicolumn{1}{c|}{$\dupD_S(0)= 2.190\ \si{ps}$} & $\dupD_S(\infty)= 3.7226 \ \si{ps}$      \\ \hline
\multicolumn{2}{|c|}{$\ddoD_S(-\infty)=3.8054 \ \si{ps}$} & \multicolumn{1}{c|}{$\ddoD_S(0)=4.107 \ \si{ps}$} & $\ddoD_S(\infty)=3.616 \ \si{ps}$      \\ \hline
\multicolumn{4}{|c|}{Parameters for $\dmin=299  \ \si{fs}$, $C=0.9431  \ \si{fF}$:}                                           \\ \hline
\multicolumn{2}{|c|}{$R_{n_{A}}=4.40882407303950 \ k \si{\ohm}$} & \multicolumn{1}{c|}{$R_{n_{B}}=3.88442001507398 \ k \si{\ohm}$} & $R_5=828.494754381781 \ \si{\ohm}$      \\ \hline
\multicolumn{2}{|c|}{$R=1.70576915128527 \ k \si{\ohm}$}         & \multicolumn{1}{c|}{$\alpha_1=966.421722237134e-12 \ \si{\ohm} s$}  & $\alpha_2=633.741820902669e-12 \ \si{\ohm} s$ \\ \hline
\end{tabular}}
\end{table}

In addition, using
the standard features of the \invt, we also determined the resulting model parameters for the
original \NOR\ gates with both standard involution channels and IDM channels (ExpChannels)
at the output, as already used in \cite{OMFS20:INTEGRATION}.

We conducted a series of simulation experiments using the Involution Tool, where
the three input signals $I_1$--$I_3$ of our \NOR\ chain circuit were stimulated by some number $N$ of transitions,
randomly and independently generated according to a normal distribution with mean $\mu$ and
standard deviation $\sigma$. For each of our choices $(\mu,\sigma)$, we
chose $N$ to ensure roughly equal average trace lengths and conducted
$30$ simulation runs for computing the average.
\cref{fig:NORPar_acc} shows the results, which are given as the improvement (in \%) of the
average total area deviation of (i) our model vs. inertial delays resp.\ (ii) of the IDM vs. inertial
delays.
The superiority of our
model over the IDM is indeed remarkable.

\cref{fig:NORPar_TC} shows the average simulation times, which (despite our
Python implementation) compare very favorably to the excessive analog
simulation times.  

\begin{figure}[h]
	\centering
	\includegraphics[width=0.55\linewidth]{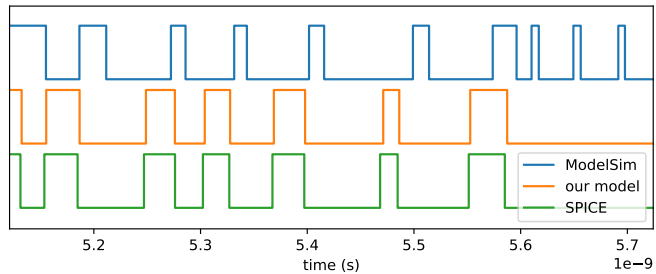}
	\caption{\scriptsize Example waveform at one output of stage 50 of our \NOR\ chain circuit.}
	\label{fig:NORchainwaveform}
\end{figure}

The superiority of our model is also confirmed by the sample waveform shown
in \cref{fig:NORchainwaveform}, which shows one of the output signals at stage 50 of our \NOR\ chain: Whereas
the trace generated from inertial delays suffers from substantial shifts and spurious glitches, the one
generated by our model faithfully matches the \spice-generated reference trace.

In order to find out whether our model is also capable of substantially
improving the somewhat modest accuracy of the IDM for simple circuits
with multi-input gates like 
\texttt{c\_17\_slack} \cite{ISCAS85_reference} (as reported 
in \cite{OMFS20:INTEGRATION}), 
we replaced all the \NAND\ gates in this circuit by \NOR\ gates
and synthesized the resulting circuit \cref{fig:c17_slack}
in the the Nangate Open Cell Library with FreePDK15$^\text{TM}$
\SI{15}{\nm} FinFET models \cite{Nangate15} ($\vdd=\SI{0.8}{\V}$) again.
Analogous to our \NOR\ chain circuit, we parametrized all its constituent gates 
individually, and performed 30 simulation runs with the same input 
stimuli $\mu=15\ \si{ps}$ and $\sigma= 5 \ \si{ps}$ as
used in \cite{OMFS20:INTEGRATION}. For $N=2000$, we obtained an
improvement of $20.8$~\% over inertial delays, which is comparable
to the improvement provided by the ExpChannel-IDM. We hence conclude
that the conjecture of the authors of the authors of 
\cite{OMFS20:INTEGRATION}, namely, that the modest accuracy
of the IDM for \texttt{c\_17\_slack} is due to its inherent inability
to handle MIS effects, is wrong. Rather, we conclude that even inertial
delay models are not as bad for simulating such simple circuits.

\section{Conclusions}
\label{Sec:con}
We have presented a complete delay modeling and simulation framework for multi-input CMOS gates that simultaneously accounts for drafting and multi-input switching effects within a fully digital dynamic timing analysis (DDTA) flow. Building on the thresholded hybrid first-order model for interconnect-augmented 2-input \NOR\ gates proposed in~\cite{ferdowsi2024hybrid}, we derived analytic delay and trajectory expressions $\dup(T,\Delta)$ and $\ddo(T,\Delta)$ that depend both on the previous-output-to-input delay $T$ and the input separation time $\Delta$. In contrast to traditional single-history channel models and static timing abstractions, the resulting model faithfully captures the subtle interaction of waveform history, pulse degradation, and MIS effects at the gate level.

A central ingredient of our development is the notion of real and virtual output transitions, which yields a one-to-one correspondence between input and output events. This abstraction enables a compact discrete-event simulation algorithm that, for each input transition, evaluates a closed-form delay formula and an associated trajectory expression to update the internal analog state. The algorithm covers all eight possible 2-input transition patterns of a \NOR\ gate, including cancellation of short pulses, and can be evaluated with constant computational effort per event. Together with the explicit characterization of the internal voltage trajectories, this provides a conceptually simple yet expressive DDTA model that scales to long chains and larger networks of such gates. On the modeling side, a key advantage of our approach lies in its parametrization procedure. Leveraging the analytic structure of the hybrid model, all relevant parameters of an interconnect-augmented \NOR\ (and, via duality, \NAND) gate can be inferred from just three characteristic MIS delay values for rising and falling transitions, along with an uncritical choice of the pure delay $\dmin$ and the extracted load capacitance. In particular, matching $\delta(0)$, $\delta(\infty)$, and $\delta(-\infty)$ for both polarities suffices to reconstruct the internal resistive and capacitive parameters, without resorting to extensive gate-level regression or numerical fitting. This substantially simplifies library characterization and makes the model attractive for integration into existing design flows.

We have implemented the resulting delay model and simulation algorithm as an extension of the Involution Tool, which now supports multi-input \NOR\ and \NAND\ gates in addition to its original IDM-based single-input channels. Thanks to the modular architecture and the existing multi-input interface, the new model could be realized as a Python-based delay engine tightly coupled to Questa’s event-driven simulation. The extended toolchain retains all features of the original framework, including random stimulus generation, parameter sweeping, detailed timing deviation metrics, and power estimation support via switching activity files. Our experiments with a deeply cascaded cross-coupled \NOR\ chain in a \SI{15}{\nm} FinFET technology show that the proposed model substantially reduces average timing error per transition compared to inertial delay channels and clearly outperforms the original IDM when MIS and drafting effects dominate, while retaining conventional digital-simulation runtimes and preserving orders-of-magnitude speedups over analog simulation. For the \texttt{c\_17\_slack} benchmark (with \NAND\ gates replaced by \NOR\ gates), the model still achieves a $20.8\%$ improvement over inertial delays, comparable to the ExpChannel-IDM. Although simple combinational circuits benefit less from the added modeling fidelity, the accuracy gains in circuits that strongly exercise drafting and multi-input effects underscore the value of our approach for asynchronous and timing-critical subsystems where waveform history significantly affects gate delays.

\bibliographystyle{plainnat}

\bibliography{cas-refs}

@techreport{NP73:spice,
    Author = {Nagel, Laurence W. and Pederson, D.O.},
    Title = {{SPICE} ({S}imulation {P}rogram with {I}ntegrated {C}ircuit {E}mphasis)},
    Institution = {EECS Department, University of California, Berkeley},
    Year = {1973},
    OPTURL = {http://www.eecs.berkeley.edu/Pubs/TechRpts/1973/22871.html},
    Number = {UCB/ERL M382}
}

@article{BDJCAVH00,
author = {M. J. Bellido-D{\'{\i}}az
	and J. Juan-Chico
	and A. J. Acosta
	and M. Valencia
	and J. L. Huertas},
title = {Logical Modelling of Delay Degradation Effect in Static {CMOS} Gates},
journal = {IEE Proceedings -- Circuits, Devices, and Systems},
volume = 147,
number = 2,
pages = {107--117},
year = 2000
}

@article{Ung71,
author = {Stephen H. Unger},
title = {Asynchronous Sequential Switching Circuits with Unrestricted Input
Changes},
journal = {IEEE Transaction on Computers},
volume = 20,
number = 12,
pages = {1437--1444},
year = 1971
}

@book{BJV06,
author = {Manuel J. Bellido-D{\'{\i}}az and Jorge Juan-Chico and Manuel Valencia},
title = {Logic-Timing Simulation and
the Degradation Delay Model},
publisher = {Imperial College Press},
address = {London},
year = 2006
}

@inproceedings{FB95,
 author = {Favalli, M. and Benini, L.},
 title = {Analysis of glitch power dissipation in {CMOS} {IC}s},
 booktitle = {Proceedings of the 1995 international symposium on Low power desig
n},
 series = {ISLPED '95},
 year = {1995},
 isbn = {0-89791-744-8},
 location = {Dana Point, California, USA},
 pages = {123--128},
 numpages = {6},
 doi = {10.1145/224081.224103},
 acmid = {224103},
 publisher = {ACM},
 address = {New York, NY, USA},
}

@ARTICLE{FNS16:ToC,
author={Matthias F\"ugger and Thomas Nowak and Ulrich Schmid},
journal={IEEE Transactions on Computers},
title={Unfaithful Glitch Propagation in Existing Binary Circuit Models},
year={2016},
volume={65},
number={3},
pages={964-978},
doi={10.1109/TC.2015.2435791},
ISSN={0018-9340},
}

@manual{Syn:CCSM,
organization={Synopsis Inc.},
title={{CCS} Timing Library Characterization Guidelines},
month={October},
year={2016},
note={Version 3.4}
}

@manual{Cad:ECSM,
organization={Cadence Design Systems},
title={Effective Current Source Model {(ECSM)} Timing and Power Specification},
month={January},
year={2015},
note={Version 2.1.2}
}

@INPROCEEDINGS{FMNNS18:DATE,
author={M. F{\"u}gger and J. Maier and R. Najvirt and T. Nowak and U. Schmid},
booktitle={2018 Design, Automation Test in Europe Conference Exhibition (DATE)},
title={A faithful binary circuit model with adversarial noise},
year={2018},
volume={},
number={},
pages={1327-1332},
doi={10.23919/DATE.2018.8342219},
ISSN={},
month={March},
url={https://ieeexplore.ieee.org/document/8342219/}
}

@inproceedings{Nangate15,
 author = {Martins, Mayler and Matos, Jody Maick and Ribas, Renato P. and Reis, Andr{\'e} and Schlinker, Guilherme and Rech, Lucio and Michelsen, Jens},
 title = {Open Cell Library in 15Nm FreePDK Technology},
 booktitle = {Proceedings of the 2015 Symposium on International Symposium on Physical Design},
 series = {ISPD '15},
 year = {2015},
 isbn = {978-1-4503-3399-3},
 location = {Monterey, California, USA},
 pages = {171--178},
 numpages = {8},
 url = {http://doi.acm.org/10.1145/2717764.2717783},
 doi = {10.1145/2717764.2717783},
 acmid = {2717783},
 publisher = {ACM},
 address = {New York, NY, USA},
}

@INPROCEEDINGS{CharlieEffect,
author={A. J. Winstanley and A. Garivier and M. R. Greenstreet},
booktitle={Proceedings of the 8th International Symposium on Asynchronous Circuits and Systems (ASYNC)},
title={{An Event Spacing Experiment}},
year={2002},
pages={47-56}, 
doi={10.1109/ASYNC.2002.1000295},
ISSN={1522-8681},
month={April},}

@ARTICLE{FNNS19:TCAD,
author={M. {Függer} and R. {Najvirt} and T. {Nowak} and U. {Schmid}},
journal={IEEE Transactions on Computer-Aided Design of Integrated Circuits and Systems},
title={A Faithful Binary Circuit Model},
year={2020},
volume={39},
number={10},
pages={2784-2797},
doi={10.1109/TCAD.2019.2937748},
ISSN={0278-0070},}

@article{OMFS20:INTEGRATION,
title = "The Involution Tool for Accurate Digital Timing and Power Analysis",
journal = "Integration",
volume = "76",
pages = "87--98",
year = "2021",
issn = "0167-9260",
doi = "https://doi.org/10.1016/j.vlsi.2020.09.007",
author = "Daniel Öhlinger and Jürgen Maier and Matthias Függer and Ulrich Schmid",
}

@INPROCEEDINGS{CGB01:DAC,
author={Liang-Chi Chen and S. K. Gupta and M. A. Breuer},
booktitle={Proceedings of the 38th Design Automation Conference},
title={A new gate delay model for simultaneous switching and its applications},
year={2001},
pages={289-294}, 
doi={10.1109/DAC.2001.156153},
ISSN={0738-100X},
month={},}

@INPROCEEDINGS{SKJPC09:ISOCC,
author={J. Shin and J. Kim and N. Jang and E. Park and Y. Choi},
booktitle={2009 International SoC Design Conference (ISOCC)},
title={A gate delay model considering temporal proximity of Multiple Input
Switching}, 
year={2009},
pages={577-580}, 
doi={10.1109/SOCDC.2009.5423815},
month={Nov},}

@ARTICLE{RS21:TCAD,
  author={Shashank Ram, O. V. S. and Saurabh, Sneh},
  journal={IEEE Transactions on Computer-Aided Design of Integrated Circuits and Systems}, 
  title={Modeling Multiple-Input Switching in Timing Analysis Using Machine Learning}, 
  year={2021},
  volume={40},
  number={4},
  pages={723-734},
  doi={10.1109/TCAD.2020.3009624}}

@inproceedings{FMOS22:DATE,
  author    = {Arman Ferdowsi and
               J{\"{u}}rgen Maier and
               Daniel {\"{O}}hlinger and
               Ulrich Schmid},
  title     = {A Simple Hybrid Model for Accurate Delay Modeling of a Multi-Input
               Gate},
booktitle = {Proceedings of the 2022 Design, Automation \& Test in Europe Conference \& Exhibition},
  year      = {2022},
}

@ARTICLE{ShichmanHodges,
author={H. {Shichman} and D. A. {Hodges}},
journal={IEEE Journal of Solid-State Circuits},
title={Modeling and simulation of insulated-gate field-effect transistor switching circuits},
year={1968},
volume={3},
number={3},
pages={285-289},
doi={10.1109/JSSC.1968.1049902},
ISSN={0018-9200},
}

@inproceedings{ferdowsi2023accurate,
  title={Accurate Hybrid Delay Models for Dynamic Timing Analysis},
  author={Ferdowsi, Arman and Schmid, Ulrich and Salzmann, Josef},
  booktitle={2023 IEEE/ACM International Conference on Computer Aided Design (ICCAD)},
  pages={1--9},
  year={2023},
  organization={IEEE}
}

@inproceedings{CS96:DAC,
author = {Chandramouli, V. and Sakallah, Karem A.},
title = {Modeling the Effects of Temporal Proximity of Input Transitions on Gate Propagation Delay and Transition Time},
year = {1996},
isbn = {0897917790},
url = {https://doi.org/10.1145/240518.240635},
doi = {10.1145/240518.240635},
booktitle = {Proc. DAC'96},
pages = {617–622},
}

@article{najm1994survey,
  title={A survey of power estimation techniques in VLSI circuits},
  author={Najm, Farid N},
  journal={IEEE Transactions on Very Large Scale Integration (VLSI) Systems},
  volume={2},
  number={4},
  pages={446--455},
  year={1994},
  publisher={IEEE}
}

@article{ferdowsi2024faithful,
  title={Faithful Dynamic Timing Analysis of Digital Circuits Using Continuous Thresholded Mode-Switched ODEs},
  author={Ferdowsi, Arman and F{\"u}gger, Matthias and Nowak, Thomas and Schmid, Ulrich and Drmota, Michael},
  journal={arXiv preprint arXiv:2403.03235},
  year={2024}
}

@inproceedings{maier2021composable,
  title={A composable glitch-aware delay model},
  author={Maier, J{\"u}rgen and {\"O}hlinger, Daniel and Schmid, Ulrich and F{\"u}gger, Matthias and Nowak, Thomas},
  booktitle={Proceedings of the 2021 on Great Lakes Symposium on VLSI},
  pages={147--154},
  year={2021}
}

@article{ferdowsi2024hybrid,
  title={A Hybrid Delay Model for Interconnected Multi-Input Gates},
  author={Ferdowsi, Arman and F{\"u}gger, Matthias and Salzmann, Josef and Schmid, Ulrich},
  journal={arXiv preprint arXiv:2403.10540},
  year={2024}
}

@inproceedings{OS23:DDECS,
  author       = {Daniel {\"{O}}hlinger and
                  Ulrich Schmid},
  editor       = {Maksim Jenihhin and
                  Hana Kub{\'{a}}tov{\'{a}} and
                  Nele Metens and
                  Jaan Raik and
                  Foisal Ahmed and
                  Jan Belohoubek},
  title        = {A Digital Delay Model Supporting Large Adversarial Delay Variations},
  booktitle    = {26th International Symposium on Design and Diagnostics of Electronic
                  Circuits and Systems, {DDECS} 2023, Tallinn, Estonia, May 3-5, 2023},
  pages        = {111--117},
  publisher    = {{IEEE}},
  year         = {2023},
  url          = {https://doi.org/10.1109/DDECS57882.2023.10139680},
  doi          = {10.1109/DDECS57882.2023.10139680},
}

@article{ohlinger2018involution,
  title={Involution tool},
  author={{\"O}hlinger, Daniel},
  journal={E191-Institut f{\"u}r Computer Engineering},
  year={2018}
}

@Misc{Encounter_tutorial2016,
  author       = {Joao Canas Ferreira},
  title        = {Physical synthesis with Encounter (Cadence)},
  howpublished = {\url{https://paginas.fe.up.pt/~jcf/ensino/disciplinas/mieec/pcvlsi/2015-16/tut_encounter/tut_encounter.html}},
  year         = {2015/16},
}

@article{ferdowsi2025faithful,
title = {Faithful dynamic timing analysis of digital circuits using continuous thresholded mode-switched ODEs},
journal = {Nonlinear Analysis: Hybrid Systems},
volume = {56},
pages = {101572},
year = {2025},
issn = {1751-570X},
doi = {https://doi.org/10.1016/j.nahs.2024.101572},
url = {https://www.sciencedirect.com/science/article/pii/S1751570X24001092},
author = {Arman Ferdowsi and Matthias Függer and Thomas Nowak and Ulrich Schmid and Michael Drmota}
}

@article{BCSS08:TCAD,
author = {Blaauw, D. and Chopra, K. and Srivastava, A. and Scheffer, L.},
title = {Statistical Timing Analysis: From Basic Principles to State of the Art},
year = {2008},
issue_date = {April 2008},
publisher = {IEEE Press},
volume = {27},
number = {4},
issn = {0278-0070},
url = {https://doi.org/10.1109/TCAD.2007.907047},
doi = {10.1109/TCAD.2007.907047},
journal = {Trans. Comp.-Aided Des. Integ. Cir. Sys.},
month = apr,
pages = {589–607},
numpages = {19}
}

@article{FP09:VLSI,
author = {Forzan, Cristiano and Pandini, Davide},
title = {Statistical static timing analysis: A survey},
year = {2009},
issue_date = {June, 2009},
publisher = {Elsevier Science Publishers B. V.},
address = {NLD},
volume = {42},
number = {3},
issn = {0167-9260},
url = {https://doi.org/10.1016/j.vlsi.2008.10.002},
doi = {10.1016/j.vlsi.2008.10.002},
journal = {Integr. VLSI J.},
month = jun,
pages = {409–435},
numpages = {27}
}

@inproceedings{FTO08:DAC,
author = {Fukuoka, Takayuki and Tsuchiya, Akira and Onodera, Hidetoshi},
title = {Statistical gate delay model for multiple input switching},
year = {2008},
isbn = {9781424419227},
publisher = {IEEE Computer Society Press},
address = {Washington, DC, USA},
booktitle = {Proceedings of the 2008 Asia and South Pacific Design Automation Conference},
pages = {286–291},
numpages = {6},
location = {Seoul, Korea},
series = {ASP-DAC '08}
}

@INPROCEEDINGS{ADB04:DAC,
  author={Agarwal, A. and Dartu, F. and Blaauw, D.},
  booktitle={Proceedings. 41st Design Automation Conference, 2004.}, 
  title={Statistical gate delay model considering multiple input switching}, 
  year={2004},
  volume={},
  number={},
  pages={658-663},
  doi={10.1145/996566.996746}}

@inproceedings{AKMK06:DAC,
author = {Amin, Chirayu and Kashyap, Chandramouli and Menezes, Noel and Killpack, Kip and Chiprout, Eli},
title = {A multi-port current source model for multiple-input switching effects in CMOS library cells},
year = {2006},
isbn = {1595933816},
publisher = {Association for Computing Machinery},
address = {New York, NY, USA},
url = {https://doi.org/10.1145/1146909.1146974},
doi = {10.1145/1146909.1146974},
booktitle = {Proceedings of the 43rd Annual Design Automation Conference},
pages = {247–252},
numpages = {6},
location = {San Francisco, CA, USA},
series = {DAC '06}
}

@article{SRC15:TODAES,
author = {Subramaniam, Anupama R. and Roveda, Janet and Cao, Yu},
title = {A Finite-Point Method for Efficient Gate Characterization Under Multiple Input Switching},
year = {2015},
issue_date = {November 2015},
publisher = {Association for Computing Machinery},
address = {New York, NY, USA},
volume = {21},
number = {1},
issn = {1084-4309},
url = {https://doi.org/10.1145/2778970},
doi = {10.1145/2778970},
journal = {ACM Trans. Des. Autom. Electron. Syst.},
month = dec,
articleno = {10},
numpages = {25}
}

@inproceedings{SRPW20:DAC,
author = {Sinha, Debjit and Rao, Vasant and Peddawad, Chaitanya and Wood, Michael and Hemmett, Jeffrey and Skariah, Suriya and Williams, Patrick},
title = {Statistical timing analysis considering multiple-input switching},
year = {2020},
isbn = {9781450367257},
publisher = {IEEE Press},
booktitle = {Proceedings of the 57th ACM/EDAC/IEEE Design Automation Conference},
articleno = {227},
numpages = {6},
location = {Virtual Event, USA},
series = {DAC '20}
}

@INPROCEEDINGS{SZ05:ICCAD,
  author={Sinha, D. and Hai Zhou},
  booktitle={ICCAD-2005. IEEE/ACM International Conference on Computer-Aided Design, 2005.}, 
  title={A unified framework for statistical timing analysis with coupling and multiple input switching}, 
  year={2005},
  volume={},
  number={},
  pages={837-843},
  doi={10.1109/ICCAD.2005.1560179}}

@INPROCEEDINGS{TZBM11,
  author={Tang, Qin and Zjajo, Amir and Berkelaar, Michel and van der Meijs, Nick},
  booktitle={2011 IEEE International Conference on IC Design \& Technology}, 
  title={Statistical delay calculation with Multiple Input Simultaneous Switching}, 
  year={2011},
  volume={},
  number={},
  pages={1-4},
  doi={10.1109/ICICDT.2011.5783205}}

@INPROCEEDINGS{YLW05,
  author={Yanamanamanda, S. and Jun Li and Wang, J.},
  booktitle={2005 IEEE International Symposium on Circuits and Systems (ISCAS)}, 
  title={Uncertainty modeling of gate delay considering multiple input switching}, 
  year={2005},
  volume={},
  number={},
  pages={2457-2460 Vol. 3},
  doi={10.1109/ISCAS.2005.1465123}}

@ARTICLE{ISCAS85_reference,
  author={Hansen, M.C. and Yalcin, H. and Hayes, J.P.},
  journal={IEEE Design \& Test of Computers}, 
  title={Unveiling the ISCAS-85 benchmarks: a case study in reverse engineering}, 
  year={1999},
  volume={16},
  number={3},
  pages={72-80},
  doi={10.1109/54.785838}
}

@inproceedings{FFNS23:HSCC,
author = {Ferdowsi, Arman and F\"{u}gger, Matthias and Nowak, Thomas and  Schmid, Ulrich},
title = {Continuity of Thresholded Mode-Switched ODEs and Digital Circuit Delay Models},
year = {2023},
isbn = {9798400700330},
publisher = {Association for Computing Machinery},
address = {New York, NY, USA},
url = {https://doi.org/10.1145/3575870.3587125},
doi = {10.1145/3575870.3587125},
booktitle = {Proceedings of the 26th ACM International Conference on Hybrid Systems: Computation and Control},
articleno = {10},
numpages = {11},
location = {San Antonio, TX, USA},
series = {HSCC'23}
}

@CONFERENCE{CC04,
	author = {Clariso, Robert and Cortadella, Jordi},
	title = {Verification of timed circuits with symbolic delays},
	year = {2004},
	journal = {Proceedings of the Asia and South Pacific Design Automation Conference, ASP-DAC},
	pages = {628 – 633},
	type = {Conference paper},
	publication_stage = {Final},
	source = {Scopus},
}

@misc{TEFS25:arxiv,
      title={Symbolic Timing Analysis of Digital Circuits Using Analytic Delay Functions}, 
      author={Era Thaqi and Dennis Eigner and Arman Ferdowsi and Ulrich Schmid},
      year={2025},
      eprint={2510.15907},
      archivePrefix={arXiv},
      primaryClass={cs.AR},
      url={https://arxiv.org/abs/2510.15907}, 
}

@article{luinsta,
  title={INSTA: An Ultra-Fast, Differentiable, Statistical Static Timing Analysis Engine for Industrial Physical Design Applications},
  author={Lu, Yi-Chen and Guo, Zizheng and Kunal, Kishor and Liang, Rongjian and Ren, Haoxing}
}

@article{AH24,
title = {Formal timing analysis of gate-level digital circuits using model checking},
journal = {Microprocessors and Microsystems},
volume = {109},
pages = {105083},
year = {2024},
issn = {0141-9331},
doi = {https://doi.org/10.1016/j.micpro.2024.105083},
url = {https://www.sciencedirect.com/science/article/pii/S0141933124000784},
author = {{Qurat-ul} Ain and Osman Hasan}
}

\end{document}